\documentclass[journal]{IEEEtran}
\usepackage{amsmath,amssymb,bm,mathtools,multirow,arydshln,cite}
\newcommand{\E}{\ensuremath{\operatorname{E}}}
\def\H{\mathrm{\scriptscriptstyle H}}
\newcommand{\ri}{\ensuremath{\mathrm{i}}}
\begin{document}
\title{Frequency-Domain Stochastic Modeling of Stationary Bivariate or Complex-Valued Signals}
\author{Adam~M.~Sykulski,~\IEEEmembership{Member,~IEEE,}
        Sofia~C.~Olhede,~\IEEEmembership{Member,~IEEE,}
        Jonathan~M.~Lilly,~\IEEEmembership{Senior~Member,~IEEE,} and~Jeffrey~J.~Early% <-this % stops a space
         \thanks{The work of A. M. Sykulski was supported by a Marie Curie International Outgoing Fellowship within the 7th European Community Framework Programme and the UK Engineering and Physical Sciences Research Council via EP/I005250/1. S. C. Olhede acknowledges funding from the UK Engineering and Physical Sciences Research Council via EP/I005250/1 and EP/L025744/1 as well as from the European Research Council via Grant CoG 2015-682172NETS within the Seventh European Union Framework Program. The work of J. M. Lilly and J. J. Early was supported by award 1235310 from the Physical Oceanography program of the United States National Science Foundation.}
\thanks{A. M. Sykulski and S. C. Olhede are with the Department of Statistical Science, University College London, Gower Street, London WC1E 6BT, UK (emails: a.sykulski@ucl.ac.uk, s.olhede@ucl.ac.uk).}% <-this % stops a space
\thanks{J. M. Lilly and J. J. Early are with NorthWest Research Associates, PO Box 3027, Bellevue, WA, USA (emails: lilly@nwra.com, jearly@nwra.com)}% <-this % stops a space
}
%\markboth{IEEE Transactions on Signal Processing}%
\maketitle
%%%%%%%%%%%%%%%%%%%%%%%%%%%%%%%%%%%%%%%%%%%%%%%%%%%%%%%%%%%%%%%%%%%%%%%%%%%%%%%%%%%%%%%%%%%%%%%%%%%%%%%%%%%%%%%%%%%%%%%%%%%%%%%%%%%%%%%%%%%%%
\begin{abstract}
There are three equivalent ways of representing two jointly observed real-valued signals: as a bivariate vector signal, as a single complex-valued signal, or as two analytic signals known as the rotary components. Each representation has unique advantages depending on the system of interest and the application goals. In this paper we provide a joint framework for all three representations in the context of frequency-domain stochastic modeling. This framework allows us to extend many established statistical procedures for bivariate vector time series to complex-valued and rotary representations. These include procedures for parametrically modeling signal coherence, estimating model parameters using the Whittle likelihood, performing semi-parametric modeling, and choosing between classes of nested models using model choice. We also provide a new method of testing for impropriety in complex-valued signals, which tests for noncircular or anisotropic second-order statistical structure when the signal is represented in the complex plane. Finally, we demonstrate the usefulness of our methodology in capturing the anisotropic structure of signals observed from fluid dynamic simulations of turbulence.
\end{abstract}
\noindent
\copyright 2017 IEEE. Personal use of this material is permitted. Permission from IEEE must be obtained for all other uses, in any current or future media, including reprinting/republishing this material for advertising or promotional purposes, creating new collective works, for resale or redistribution to servers or lists, or reuse of any copyrighted component of this work in other works.
%%%%%%%%%%%%%%%%%%%%%%%%%%%%%%%%%%%%%%%%%%%%%%%%%%%%%%%%%%%%%%%%%%%%%%%%%%%%%%%%%%%%%%%%%%%%%%%%%%%%%%%%%%%%%%%%%
\section{Introduction}
\IEEEPARstart{I}{n} many applications of signal processing, there is a need to jointly analyze two real-valued signals when they share a common dependence structure. Examples include radio frequency position and displacement measurements of bloodflow~\cite{brands1997radio}, and eastward and northward geophysical signals such as wind or ocean current velocities~\cite{gonella1972rotary}. There are three distinct mathematical ways of representing two real-valued signals: as a bivariate vector signal, as a single complex-valued signal, and as two complex-valued analytic signals known as the rotary components. Each representation has unique advantages depending on the system of interest and the application goals.

\begin{figure*}
\centering
\includegraphics[trim={1cm 0.1cm 1.6cm 1.7cm},width=0.42\linewidth]{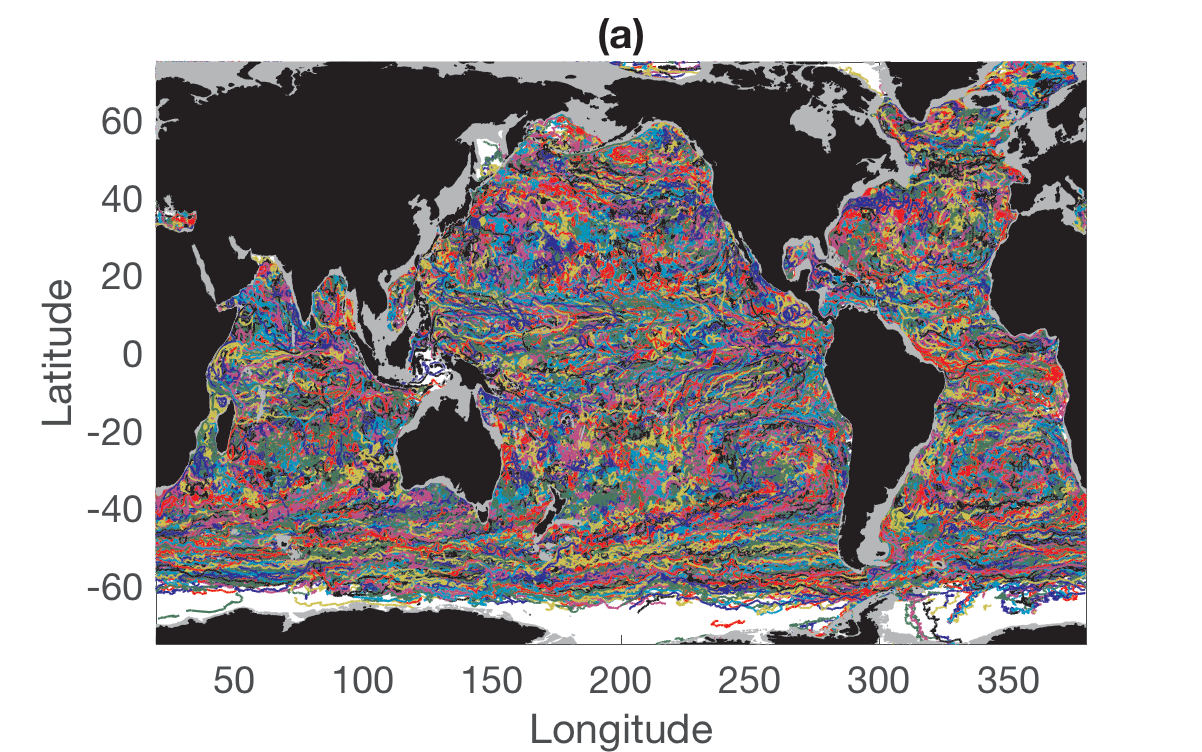}
\includegraphics[width=0.42\linewidth]{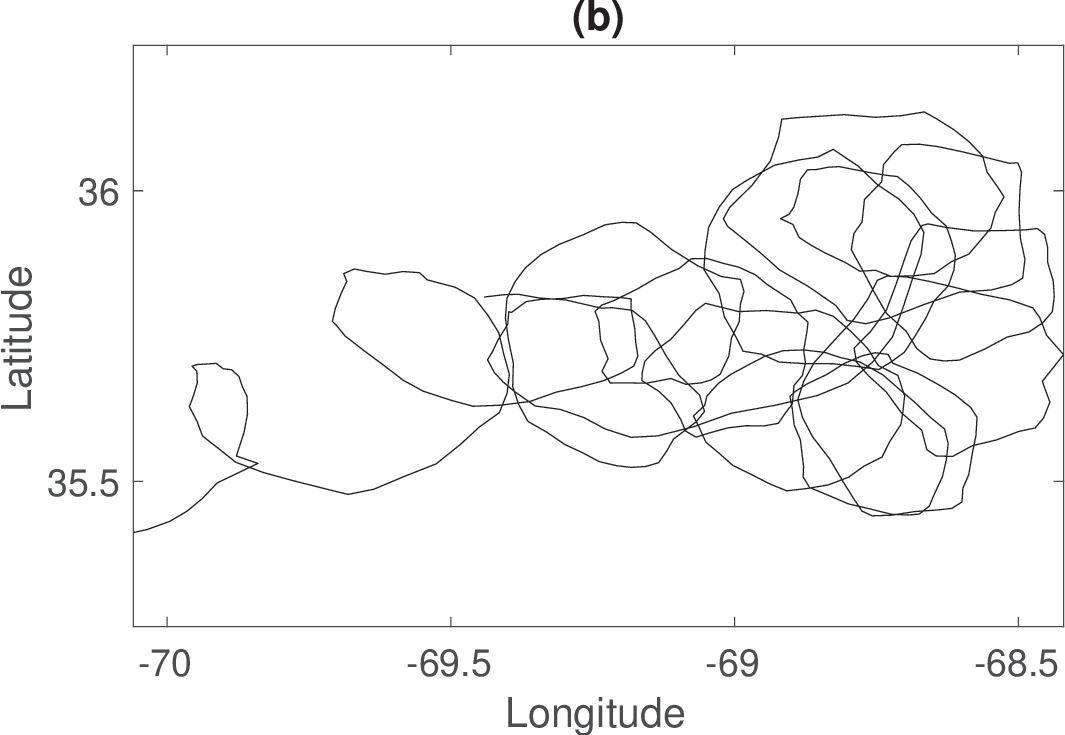}\vspace{3mm}
\includegraphics[width=0.42\linewidth]{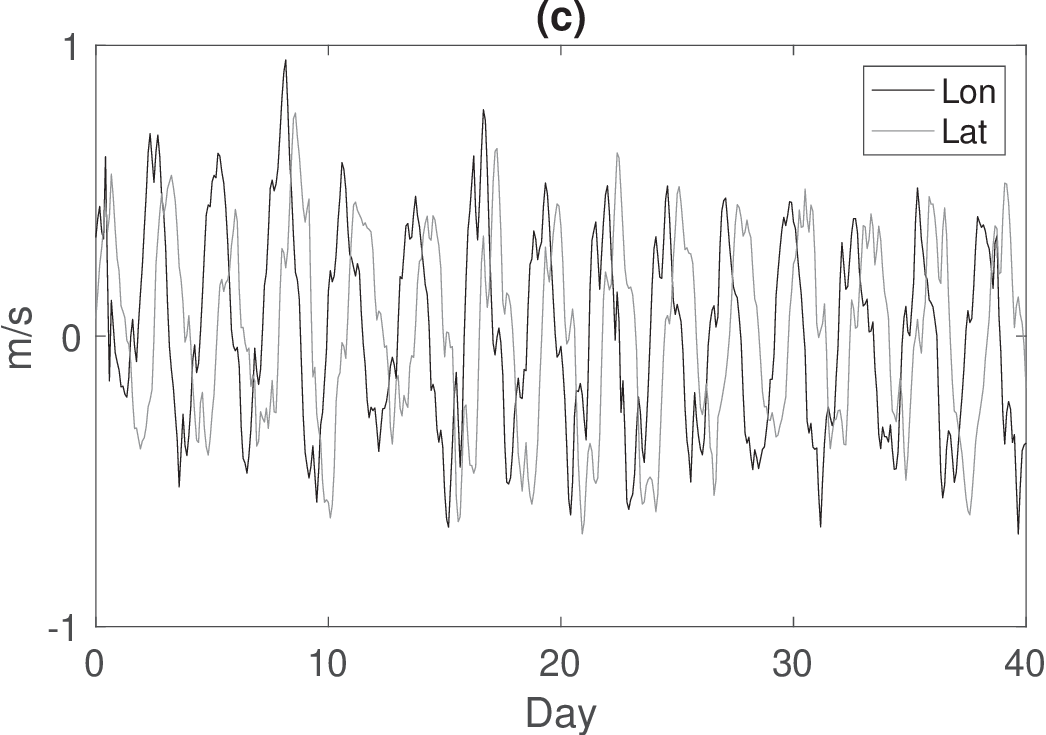}
\includegraphics[width=0.42\linewidth]{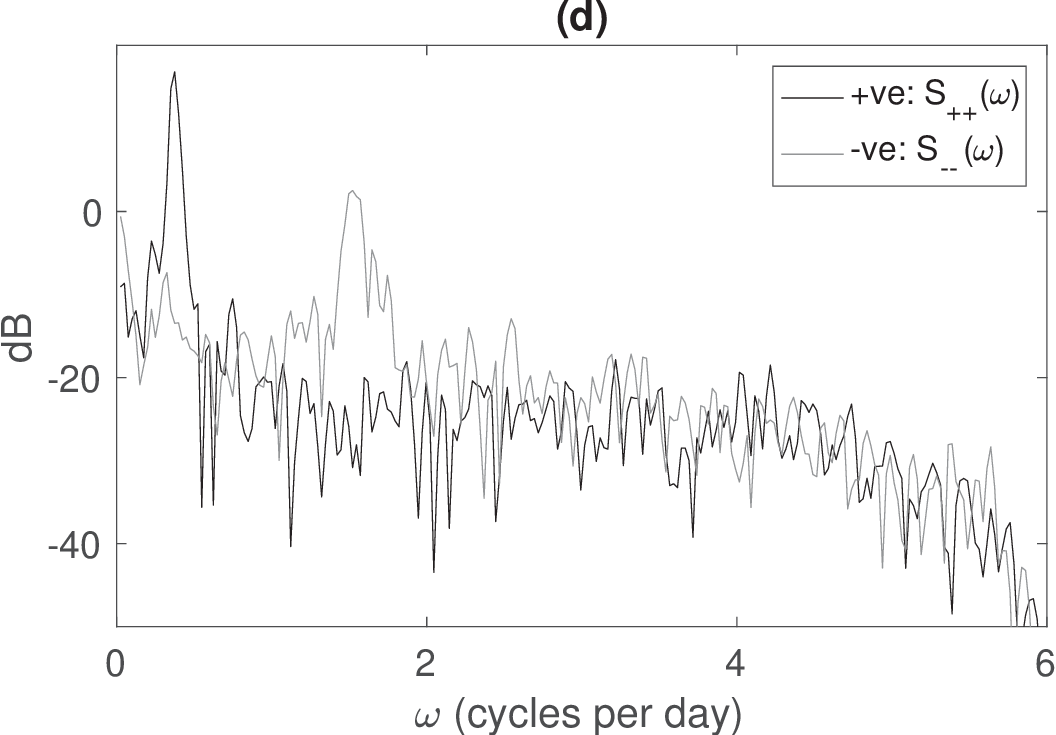}
\caption{\label{FigIntro}(a) satellite-tracked position trajectories from multiple freely-drifting instruments from the Global Drifter Program (\texttt{www.aoml.noaa.gov/phod/dac}). (b) a 40-day position trajectory of North Atlantic drifter ID\#44000. (c) velocities of this drifter over time in each Cartesian direction. (d) a multi-taper spectral density estimate of the rotary components, using the discrete prolate spheroidal sequence (dpss) tapers with bandwidth parameter 3, with negative and positive frequencies overlaid.}
\end{figure*}

To motivate the need for these different representations, in Fig.~\ref{FigIntro}(a) we plot the satellite-tracked position trajectories of a large array of freely-drifting oceanographic instruments obtained from the Global Drifter Program~\cite{lumpkin07}.
In Figs.~\ref{FigIntro}(b) and \ref{FigIntro}(c) we plot a 40-day position trajectory from a North-Atlantic drifter, together with the velocities corresponding to the eastward and northward displacements of this trajectory. In Fig.~\ref{FigIntro}(d) we display a multi-taper spectral density estimate of a complex-valued velocity signal constructed from Fig.~\ref{FigIntro}(c). The spectrum is supported over both negative and positive frequencies, distinguishing oscillatory behavior with a preferred direction of rotation, known commonly as the rotary components~\cite{gonella1972rotary}. Fig.~\ref{FigIntro}(d) reveals that the signal contains two counter-rotating oscillations at different frequencies. This is not as easily observed in the bivariate time-domain representation of Fig.~\ref{FigIntro}(c), thus motivating the benefits of considering different representations of two jointly observed signals.

In many applications there is a need to specify a simple parametric model for the signal structure, and then to estimate these parameters from a set of observed signals. In this paper we describe a framework for parametrically modeling and estimating the parameters of two real-valued signals in each of the three representations as stationary Gaussian stochastic processes. This builds on ideas found in \cite{walker1993complex,mandic2009complex,schreier2010statistical,lilly2010bivariate,walden2013rotary} for non-parametric or deterministic modeling of complex-valued signals, with further understanding developed in \cite{davenport1961spectrum,gonella1972rotary,calman1978interpretation} for atmospheric and oceanographic processes.

Approaches to stochastic parametric modeling of complex-valued signals have been primarily focused on the class of autoregressive moving average (ARMA) models using widely-linear filters, see e.g.~\cite{picinbono1997second,rubin2008kinematics,navarro2008arma,sykulski2015improper}.
In this paper, we propose more general classes of stochastic parametric models for complex-valued and rotary representations; and we connect these with well-known bivariate modeling techniques such as~\cite{hamon1974spectral}. 
The background for mathematically connecting the representations is given in Section~\ref{S:prelim}. Then in~Section~\ref{S:Models} we propose novel parametric models for specifying either the bivariate coherency or the rotary coherency---two important quantities which are shown to be distinct, thus providing new and easily interpreted model structures.

Our joint framework allows well-known procedures for bivariate modeling to be extended to complex-valued and rotary representations. First in Section~\ref{S:Whittle} we provide computationally efficient procedures for parameter estimation, by extending the Whittle likelihood from the bivariate to complex-valued and rotary representations. Then in Section~\ref{semipm} we demonstrate how the Whittle likelihood can naturally be used with semi-parametric modeling techniques (relating this to the seminal work of~\cite{robinson1995gaussian}). In Section~\ref{impropriety} we construct a parametric test for impropriety, which also tests for anisotropy if the signals are tracking spatial positions like the ocean drifters. In Section~\ref{modelchoice}, we detail how model choice is correctly performed with complex-valued and rotary signals when selecting between nested models. Finally, in Section \ref{S:Simulations} we demonstrate the usefulness of our methodology in developing key physical understanding, by applying our methods to signals obtained from fluid dynamic models of turbulence.
%%%%%%%%%%%%%%%%%%%%%%%%%%%%%%%%%%%%%%%%%%%%%%%%%%%
%%%%%%%%%%%%%%%%%%%%%%%%%%%%%%%%%%%%%%%%%%%%%%%%%%%
\section{Background}\label{S:prelim}
%%%%%%%%%%%%%%%%%%%%%%%%%%%%%%%%%%%%%%%%%%%%%%%%%%%
A bivariate pair of real-valued signals can be given two other equivalent representations: as a complex-valued signal, or as a pair of analytic signals which we call the rotary components. The information contained in each of these three representations---bivariate/complex/rotary---is equivalent, but the operations required to transform between them are in general nontrivial. In this section we provide a brief background with necessary definitions and notation, culminating with Table~\ref{Table:RealComplex} which links the various spectral representations. Much of the material in this section can be found in~\cite{walker1993complex,mandic2009complex,schreier2010statistical,lilly2010bivariate,walden2013rotary}.

\subsection{Bivariate Processes}
Consider a zero-mean continuous-time bivariate Gaussian process, denoted at time $t$ by $[X(t)\;Y(t)]^T$, where ``$T$'' denotes matrix transpose. Using the Cram\'er spectral representation theorem, we define this process in terms of the orthogonal increments processes $d\Psi_X(\omega)$ and $d\Psi_Y(\omega)$ such that
\begin{equation}
\begin{bmatrix}
X(t) \\ Y(t)
\end{bmatrix}
\equiv \frac{1}{2\pi}
\int
\begin{bmatrix}
d\Psi_X(\omega) \\
d\Psi_Y(\omega)
\end{bmatrix}
e^{\ri\omega t},
\label{spectral_rep}
\end{equation}
where $\ri\equiv\sqrt{-1}$. The statistics of the bivariate process can also be fully specified by the power spectral matrix $\bm{S}_B(\omega)$ defined by
\begin{multline}
\bm{S}_B(\omega)\delta(\omega-\nu)d\omega d\nu \equiv \\ \frac{1}{2\pi} \E \left\{\begin{bmatrix}d\Psi_X(\omega) \\d\Psi_Y(\omega)\end{bmatrix}\begin{bmatrix}d\Psi_X^\ast(\nu) & d\Psi_Y^\ast(\nu)\end{bmatrix}\right\},
\label{eq:CramerB}
\end{multline}
where $\delta(\cdot)$ is the Dirac-delta function and $\ast$ denotes the complex conjugate. We denote the elements of $\bm{S}_B(\omega)$ by
\begin{equation}\label{eq:SB}
\bm{S}_B(\omega)=\begin{bmatrix}
S_{XX}(\omega) & S_{XY}(\omega)\\
S_{XY}^\ast(\omega) & S_{YY}(\omega)
\end{bmatrix}.
\end{equation}
The power spectral densities $S_{XX}(\omega)$ and $S_{YY}(\omega)$ are real-valued, nonnegative, and symmetric in $\omega$, whereas the cross-spectral density $S_{XY}(\omega)$ is complex-valued and Hermitian symmetric. Finally, the statistics of the bivariate process can be fully specified by the covariance matrix $\bm{R}_B(\tau)$ defined by
\begin{equation}
\bm{R}_B(\tau) \equiv \E \left\{\begin{bmatrix}X(t) \\ Y(t)\end{bmatrix}\begin{bmatrix}X(t-\tau) & Y(t-\tau)\end{bmatrix}\right\},
\end{equation}
where the elements of the matrix $\bm{R}_B(\tau)$ are denoted by
\begin{equation}\label{eq:RU}
\bm{R}_B(\tau)=\begin{bmatrix}
s_{XX}(\tau) & s_{XY}(\tau)\\
s_{XY}^\ast(\tau) & s_{YY}(\tau)
\end{bmatrix}.
\end{equation}
From these definitions it follows that
\begin{multline}
\bm{R}_B(\tau) = \\ \frac{1}{4\pi^2}\int\int e^{\ri\omega t} e^{-\ri\nu (t-\tau)}\E \left\{\begin{bmatrix}d\Psi_X(\omega) \\d\Psi_Y(\omega)\end{bmatrix}\begin{bmatrix}d\Psi_X^\ast(\nu) & d\Psi_Y^\ast(\nu)\end{bmatrix}\right\} \\ = \frac{1}{2\pi}\int\bm{S}_B(\omega)e^{i\omega \tau} d\omega,
\label{eq:fourier_pair}
\end{multline}
such that the covariance matrix $\bm{R}_B(\tau)$ forms a Fourier pair with the power spectral matrix $\bm{S}_B(\omega)$.

\begin{table*}
\caption{\label{Table:RealComplex}Relationships between spectra of three different representations of two jointly observed real-valued signals: Bivariate (left column), complex-valued (middle column), and rotary (right column)}
\centering
\renewcommand{\arraystretch}{1.4}
\begin{tabular}{c|c|c|c}\noindent
& Bivariate ($X/Y$) & Complex ($Z/Z^\ast$) & Rotary ($Z^+/Z^-$) $(\omega\neq0)$ \\ \cline{1-4}
\multirow{6}{*}{\rotatebox[origin=c]{90}{Bivariate}} & \multirow{2}{*}{$S_{XX}(\omega)$} & $\frac{1}{4}\left[S_{ZZ}(\omega)+S_{ZZ}(-\omega)\right]$ &
$\frac{1}{4}\left[S_{++}(|\omega|)+S_{--}(|\omega|)\right]$ \\ & & $+\frac{1}{2}\Re\{R_{ZZ}(\omega)\}$ & $+\frac{1}{2}\Re\{S_{+-}(|\omega|\}$ \\ \cdashline{2-4}
& \multirow{2}{*}{$S_{YY}(\omega)$} & $\frac{1}{4}\left[S_{ZZ}(\omega)+S_{ZZ}(-\omega)\right]$ & $\frac{1}{4}\left[S_{++}(|\omega|)+S_{--}(|\omega|)\right]$\\ & & $-\frac{1}{2}\Re\{R_{ZZ}(\omega)\}$ & $-\frac{1}{2}\Re\{S_{+-}(|\omega|\}$ \\ \cdashline{2-4}
& \multirow{2}{*}{$S_{XY}(\omega)$} & $\frac{1}{2}\Im\left\{R_{ZZ}(\omega)\right\}$ & 
$\frac{1}{2}\Im\left\{S_{+-}(|\omega|)\right\}$\\
& & $+\frac{1}{4}\ri\left[S_{ZZ}(\omega)-S_{ZZ}(-\omega)\right]$ & $+\frac{\ri}{4}\mathrm{sgn}(\omega)\left[S_{++}(|\omega|)-S_{--}(|\omega|)\right]$\\ \cline{1-4}
\multirow{4}{*}{\rotatebox[origin=c]{90}{Complex}} & $S_{XX}(\omega)+S_{YY}(\omega)$ & \multirow{2}{*}{$S_{ZZ}(\omega)$} & \multirow{2}{*}{$S_{++}(\omega)+S_{--}(-\omega)$}\\
& $+2\Im\left\{S_{XY}(\omega)\right\}$ & & \\ \cdashline{2-4}
& $S_{XX}(\omega)-S_{YY}(\omega)$ & \multirow{2}{*}{$R_{ZZ}(\omega)$} & \multirow{2}{*}{$S_{+-}(\omega)+S_{+-}(-\omega)$}\\
& $+2\ri\Re\left\{S_{XY}(\omega)\right\}$ & & \\ \cline{1-4}
\multirow{6}{*}{\rotatebox[origin=c]{90}{Rotary $(\omega>0)$}} & $S_{XX}(\omega)+S_{YY}(\omega)$ & \multirow{2}{*}{$S_{ZZ}(\omega)$} & \multirow{2}{*}{$S_{++}(\omega)$} \\ & $+2\Im\left\{S_{XY}(\omega)\right\}$ & & \\ \cdashline{2-4}
& $S_{XX}(\omega)+S_{YY}(\omega)$ & \multirow{2}{*}{$S_{ZZ}(-\omega)$} & \multirow{2}{*}{$S_{--}(\omega)$} \\ & $-2\Im\left\{S_{XY}(\omega)\right\}$ & & \\ \cdashline{2-4}
& $S_{XX}(\omega)-S_{YY}(\omega)$ & \multirow{2}{*}{$R_{ZZ}(\omega)$} & \multirow{2}{*}{$S_{+-}(\omega)$} \\
& $+2\ri\Re\left\{S_{XY}(\omega)\right\}$ & &
\end{tabular}
\\ \vspace{2mm} The three elements in a given row are equal for the specified range of $\omega$, for example the last row reads: $S_{XX}(\omega)-S_{YY}(\omega)$ $+2\ri\Re\left\{S_{XY}(\omega)\right\}=R_{ZZ}(\omega)=S_{+-}(\omega)$ for $\omega>0$, where $\Re{\{\cdot\}}$ and $\Im{\{\cdot\}}$ refer to the real and imaginary parts respectively.
\end{table*}

\subsection{Complex-valued Processes}
The bivariate real-valued process $[X(t)\;Y(t)]^T$ can alternatively be expressed as a single complex-valued signal defined by
\begin{equation}
Z(t)\equiv X(t)+\ri Y(t).
\end{equation}
In this case the second-order statistics of $[X(t)\;Y(t)]^T$ are equivalent to the statistics of $Z(t)$ together with its complex conjugate $Z^\ast(t)$, as we now show; see also \cite{mandic2009complex,schreier2010statistical}.

The process $Z(t)$ can be expressed in terms of the orthogonal increments process $d\Psi_Z(\omega)\equiv d\Psi_X(\omega)+\ri d\Psi_Y(\omega)$ such that
\begin{equation}
Z(t) = \frac{1}{2\pi}\int d\Psi_Z(\omega)e^{\ri\omega t}.
\label{eq:orthZ}
\end{equation}
The statistics of the complex-valued process $Z(t)$ can also be fully specified from the power spectral matrix $\bm{S}_C(\omega)$ defined by
\begin{multline}
\bm{S}_C(\omega)\delta(\omega-\nu)d\omega d\nu \equiv \\ \frac{1}{2\pi} \E \left\{\begin{bmatrix}d\Psi_Z(\omega) \\d\Psi_Z^\ast(-\omega)\end{bmatrix}\begin{bmatrix}d\Psi_Z^\ast(\nu) & d\Psi_Z(-\nu)\end{bmatrix}\right\},
\label{eq:CramerC}
\end{multline}
where the elements of the matrix $\bm{S}_C(\omega)$ are denoted by
\begin{equation}\label{eq:SC}
\bm{S}_C(\omega)=\begin{bmatrix}
S_{ZZ}(\omega) &R_{ZZ}(\omega) \\
R_{ZZ}^\ast(\omega) & S_{ZZ}(-\omega) 
\end{bmatrix}.
\end{equation}
The power spectral density $S_{ZZ}(\omega)$ is real-valued and nonnegative, but not necessarily symmetric in $\omega$, whereas the complementary spectrum\footnote{We clarify a distinction in our notation whereby $R(\omega)$ refers to a complementary spectrum and $\bm{R}(\tau)$ refers to a covariance matrix. This is chosen as such to be consistent with the conventional notation in the literature.} $R_{ZZ}(\omega)$ is in general complex-valued and is always symmetric in $\omega$, i.e. $R_{ZZ}(\omega)=R_{ZZ}(-\omega)$. 

The statistics of the complex-valued process $Z(t)$ can also be fully specified by the covariance matrix $\bm{R}_C(\tau)$ defined by
\begin{equation}
\bm{R}_C(\tau) \equiv \E \left\{\begin{bmatrix}Z(t) \\ Z^\ast(t)\end{bmatrix}\begin{bmatrix}Z^\ast(t-\tau) & Z(t-\tau)\end{bmatrix}\right\},
\end{equation}
where
\begin{equation}\label{eq:RC}
\bm{R}_C(\tau)=\begin{bmatrix}
s_{ZZ}(\tau) & r_{ZZ}(\tau)\\
r_{ZZ}^\ast(\tau) & s_{ZZ}(-\tau)
\end{bmatrix}.
\end{equation}
Similarly to~\eqref{eq:fourier_pair} it can be shown that $\bm{R}_C(\tau)$ and $\bm{S}_C(\omega)$ form a Fourier pair.
If the off-diagonal term of $\bm{R}_C(\tau)$ vanishes (that is $r_{ZZ}(\tau)=0$ for all $\tau$), or equivalently if the off-diagonal term of $\bm{S}_C(\omega)$ vanishes (that is $R_{ZZ}(\omega)=0$ for all $\omega$), then the process is said to be \textit{proper} or \textit{circular}, and otherwise it is \textit{improper} or \textit{noncircular}~\cite{schreier2010statistical}. 

To link the bivariate and complex representations, we express $\bm{S}_C(\omega)$ in terms of $\bm{S}_B(\omega)$ in~\eqref{eq:SB}. To do this we first define the unitary matrix
\begin{equation}
\bm{T}\equiv\frac{1}{\sqrt{2}}\begin{bmatrix} 1 & \ri \\ 1 & -\ri
\end{bmatrix},
\label{eq:unitary}
\end{equation}
which has the property that $\bm{T}\bm{T}^\H=\bm{I}$, where $\bm{I}$ is the $2\times2$ identity matrix and $\mathrm{\scriptstyle H}$ denotes the Hermitian transpose. It follows that
\begin{equation}
\begin{bmatrix} Z(t) \\ Z^\ast(t)
\end{bmatrix}=\sqrt{2}\bm{T}\begin{bmatrix} X(t) \\ Y(t)
\end{bmatrix}.
\end{equation}
Then from~\eqref{eq:CramerC}, and using the property $d\Psi_{Z^\ast}(\omega)=d\Psi_Z^\ast(-\omega)$, where $d\Psi_{Z^\ast}(\omega)$ is defined from~\eqref{eq:orthZ} as the orthogonal increments of $Z^\ast(t)$, we have that
\begin{multline}
\bm{S}_C(\omega)\delta(\omega-\nu)d\omega d\nu \\ = \frac{1}{2\pi} \E \left\{\begin{bmatrix}d\Psi_Z(\omega) \\d\Psi_{Z^\ast}(\omega)\end{bmatrix}\begin{bmatrix}d\Psi_Z^\ast(\nu) & d\Psi_{Z^\ast}^\ast(\nu)\end{bmatrix}\right\}\\=\frac{1}{2\pi} \E \left\{\sqrt{2}\bm{T}\begin{bmatrix}d\Psi_X(\omega) \\d\Psi_Y(\omega)\end{bmatrix}\begin{bmatrix}d\Psi_X^\ast(\nu) & d\Psi_Y^\ast(\nu)\end{bmatrix}\sqrt{2}\bm{T}^\H\right\},
\end{multline}
such that from~\eqref{eq:CramerB} we can observe that
\begin{equation}
\bm{S}_C(\omega) = 2\bm{T}\bm{S}_B(\omega)\bm{T}^\H.
\label{eq:B2C}
\end{equation}
If we expand the matrices in~\eqref{eq:B2C} then we arrive at direct transformations between $\{S_{XX}(\omega),S_{YY}(\omega),S_{XY}(\omega)\}$ and $\{S_{ZZ}(\omega),R_{ZZ}(\omega)\}$, which are displayed as part of Table~\ref{Table:RealComplex}---a table which provides transformations between the power spectra for the bivariate, complex, and rotary representations.

\subsection{Rotary Components}
A third representation we consider is to split the complex-valued signal into analytic and anti-analytic components, as is commonly performed in signal processing~\cite{gabor1946theory,cohen1995time,lilly2010bivariate}. These components are also known as the rotary components~\cite{gonella1972rotary,walden2013rotary}, and we define these by
\begin{equation}
\begin{bmatrix}
Z^+(t) \\ Z^-(t)
\end{bmatrix}
\equiv \frac{1}{2\pi}
\int
\begin{bmatrix}
d\Psi_{Z^+}(\omega) \\
d\Psi_{Z^-}(\omega)
\end{bmatrix}
e^{\ri\omega t},
\end{equation}
where the orthogonal increments are given by
\begin{equation}
\begin{bmatrix}
d\Psi_{Z^+}(\omega) \\
d\Psi_{Z^-}(\omega)
\end{bmatrix} \equiv U(\omega)
\begin{bmatrix}
d\Psi_{Z}(\omega) \\
d\Psi_{Z}^\ast(-\omega)
\end{bmatrix}.
\label{eq:ComRot}
\end{equation}
Here $U(\omega)$ is the unit (or Heaviside) step function defined by
\begin{equation}
U(\omega)\equiv \left\{ \begin{array}{l l}
1 & \quad \omega>0
\\ 1/2 & \quad \omega=0
\\ 0 & \quad \omega<0.
\end{array} \right.
\label{eq:unitstep}
\end{equation}
It follows from these definitions that
\begin{equation}
Z(t)=[Z^-(t)]^\ast+Z^+(t).
\end{equation}
As in~\cite{lilly2010bivariate}, we construct the processes
$Z^+(t)$ and $Z^-(t)$ such that they are both {\em analytic}, that is, supported only on positive frequencies. The anti-analytic component of $Z(t)$ is recovered by taking the conjugate of $Z^-(t)$. Therefore $Z^+(t)$ and $[{Z^-}(t)]^\ast$ are the analytic and anti-analytic signals associated with $Z(t)$, while $Z^-(t)$ is referred to as the conjugate analytic signal. The statistics of the process can also be specified in terms of the power spectral matrix $\bm{S}_\pm(\omega)$ where 
\begin{multline}
\bm{S}_\pm(\omega)\delta(\omega-\nu)d\omega d\nu \equiv \\ \frac{1}{2\pi} \E \left\{\begin{bmatrix}d\Psi_{Z^+}(\omega) \\d\Psi_{Z^-}(\omega)\end{bmatrix}\begin{bmatrix}d\Psi_{Z^+}^\ast(\nu) & d\Psi_{Z^-}^\ast(\nu)\end{bmatrix}\right\},
\label{eq:CramerPM}
\end{multline}
where the elements of the matrix $\bm{S}_\pm(\omega)$ are denoted by
\begin{equation}\label{eq:SPM}
\bm{S}_\pm(\omega)=\begin{bmatrix}
S_{++}(\omega) &S_{+-}(\omega) \\
S_{+-}^\ast(\omega) & S_{--}(\omega) 
\end{bmatrix}.
\end{equation}
It then follows by substituting~\eqref{eq:ComRot} into~\eqref{eq:CramerPM} that 
\begin{equation}
\bm{S}_\pm(\omega) = U^2(\omega)\bm{S}_C(\omega).
\label{eq:C2PM}
\end{equation}
This then provides a direct mapping between $\{S_{++}(\omega),S_{--}(\omega),S_{+-}(\omega)\}$ and $\{S_{ZZ}(\omega),R_{ZZ}(\omega)\}$. Furthermore, by combining~\eqref{eq:B2C} and \eqref{eq:C2PM} we also arrive at transformations between $\{S_{XX}(\omega),S_{YY}(\omega),S_{XY}(\omega)\}$ and $\{S_{++}(\omega),S_{--}(\omega),S_{+-}(\omega)\}$. These are provided in Table~\ref{Table:RealComplex} for $\omega>0$. Our choice to make $Z^-(t)$ analytic simplifies the representations, as the rotary components are defined only for nonnegative frequencies, such that all the spectral densities in $\bm{S}_\pm(\omega)$ are zero for negative frequencies. The relationships between the bivariate and complex/rotary representations are nontrivial, and highlight some of the differences between working with bivariate and complex-valued data. We will refer to Table~\ref{Table:RealComplex} in future sections to provide more intuition and simplify calculations.
%%%%%%%%%%%%%%%%%%%%%%%%%%%%%%%%%%%%%%%%%%%%%%%%%%%
%%%%%%%%%%%%%%%%%%%%%%%%%%%%%%%%%%%%%%%%%%%%%%%%%%%
\section{Stochastic Modeling of Coherency}\label{S:Models}
We now introduce a framework for the specification of stochastic models for pairs of real-valued signals. The approach we shall adopt is that first a representation of bivariate/complex/rotary is selected, and then the main diagonal of the corresponding spectral matrix $\bm{S}(\omega)$ in either~\eqref{eq:SB}, \eqref{eq:SC}, or~\eqref{eq:SPM} is specified. Then the respective off-diagonal term must also be specified, which becomes equivalent to modeling the {\em coherency} of the signal. First, in Section~\ref{prelim-corr}, we propose models for the {\em bivariate coherency} in the bivariate representation; then, in Section~\ref{prelim-rotary}, we propose models for the {\em rotary coherency} in the rotary representation. Note that we do not have a subsection for modeling in the complex-valued  representation, as equivalent models can be easily constructed from the rotary representation models and using the transformations of Table~\ref{Table:RealComplex}.
%%%%%%%%%%%%%%%%%%%%%%%%%%%%%%%%%%%%%%%%%%%%%%%%%%%
\subsection{Bivariate coherency} \label{prelim-corr}
For bivariate processes, we refer to the off-diagonal entry $S_{XY}(\omega)$ of $\bm{S}_B(\omega)$ in~\eqref{eq:SB} as the bivariate cross-spectrum, and represent this object as a product of terms
\begin{align}
S_{XY}(\omega)&=\rho_{XY}(\omega)\left[S_{XX}(\omega)S_{YY}(\omega)\right]^{1/2}\nonumber\\
\rho_{XY}(\omega)&=\sigma_{XY}(\omega)e^{-\ri\theta_{XY}(\omega)},
\label{eq:cart_coherency}
\end{align}
where $|\rho_{XY}(\omega)| \le 1$  is the {\em coherency} of $X(t)$ and $Y(t)$, $\sigma_{XY}(\omega)$ is the {\em coherence}, and $\theta_{XY}(\omega)$ is the {\em group delay}, quantifying whether $X(t)$ or $Y(t)$ are leading or lagging in time at frequency-cycle $\omega$. We refer
to $\rho_{XY}(\omega)$ as the {\em bivariate coherency}. Note that $\rho_{XY}(\omega)$ is in general complex-valued, whereas the coherence $\sigma_{XY}(\omega)$ and group delay $\theta_{XY}(\omega)$ are real-valued. The coherence and group delay must satisfy
\begin{align}
\sigma_{XY}(\omega)&=\sigma_{XY}(-\omega),\quad|\sigma_{XY}(\omega)|\le1\nonumber\\
\theta_{XY}(-\omega)&=-\theta_{XY}(\omega),
\label{amp-phase}
\end{align}
on account of the required Hermitian symmetry of $S_{XY}(\omega)$. This follows because while its Fourier pair $s_{XY}(\tau)$ does not satisfy the evenness of $s_{XX}(\tau)$ or $s_{YY}(\tau)$, it is still real-valued.

From a modeling standpoint, see for example~\cite{hamon1974spectral}, we may also regard~\eqref{eq:cart_coherency} as {\em specifying} the bivariate cross-spectrum for a given choice of the individual spectra $S_{XX}(\omega)$ and $S_{YY}(\omega)$. The proposed parametric model for $X(t)$ and $Y(t)$ is a valid Gaussian process if:
\begin{enumerate}
\item $S_{XX}(\omega)\ge 0$ and $S_{YY}(\omega)\ge 0$ for all $\omega$,
\item $|\sigma_{XY}(\omega)| \le 1$ for all $\omega$, such that 1) and 2) together ensure the determinant of the spectral matrix $\bm{S}_B(\omega)$ defined in~\eqref{eq:SB}
is nonnegative for all $\omega$ (such that the spectral matrix itself is nonnegative definite),
\item $\sigma_{XY}(\omega)$ is an even function in $\omega$ and $\theta_{XY}(\omega)$ is an odd function in $\omega$ (such that $S_{XY}(\omega)$ is Hermitian symmetric), and
\item the spectral matrix $\bm{S}_B(\omega)$ is integrable.
\end{enumerate}

Given the large flexibility for specifying $\sigma_{XY}(\omega)$ and $\theta_{XY}(\omega)$, we propose some practical forms. In essence the parameter $\sigma_{XY}(\omega)$
is providing the magnitude of correlation, and $\theta_{XY}(\omega)$ is strongly linked to 
time shifts or misalignments between $X(t)$ and $Y(t)$. If the system is dispersive~\cite{hamon1974spectral} we would expect $\theta_{XY}(\omega)$ to change non-linearly in form across frequencies; if the system is non-dispersive, $\theta_{XY}(\omega)$ would be linear and the coefficient of the linear term would quantify a time-delay.

Many signals are correlated at low frequencies, but become less so at higher frequencies where erratic variability is found. We therefore would commonly find 
$\sigma_{XY}(\omega)$ to be a decaying function bounded above by unity. Three types of decay might be expected: 
\begin{enumerate}
\item compactly supported $\sigma_{XY}(\omega)$ showing no correlation magnitude above frequency $\omega_0$,
\item exponentially decaying $\sigma_{XY}(\omega)$ showing a rapid decay, \item $\sigma_{XY}(\omega)$ exhibiting slow polynomial decay. 
\end{enumerate}

A realistic parametric model for the first case, that is compactly supported coherency, would correspond to
\begin{equation}
\label{eq:sigma0}
\sigma_{XY}(\omega)=\begin{cases}
\left|\sum_{j=0}^{J} a_j \omega^{2j} \right|, & 0\leq |\omega| \leq \omega_0,\\
0, &\omega_0<|\omega|.
\end{cases}
\end{equation}
This form is straightforward and interpretable; although we must ensure that the $a_j$ coefficients are selected such that $\sigma_{XY}(\omega)$ is bounded above by unity for all $|\omega|\leq\omega_0$, and that $\sigma_{XY}(\omega_0)=0$ to achieve continuity in frequency. This model proposes no coherence for frequencies higher than $\omega_0$, and so this parameter emerges as specifying a natural timescale of the bivariate process. 

In the second case, where we allow $\omega$ to go unbounded. A convenient form for a decaying $\sigma_{XY}(\omega)$ is the logistic function given by
\begin{align}
\label{eq:sigma}
\sigma_{XY}(\omega)&=\frac{\sigma_0\left[1+e^{q(0)} \right]}{1+e^{q(\omega)}},\nonumber\\
q(\omega)&=q_0+q_1 \omega^2+\dots q_r \omega^{2r}.
\end{align}
This model guarantees that $\sigma_{XY}(\omega)$ is an even function as required in~\eqref{amp-phase}. We require that the highest-order polynomial coefficient is positive, i.e. $q_r>0$, to ensure $\lim_{\omega\rightarrow\infty} \sigma_{XY}(\pm\omega) = 0$, and the $q_i$ coefficients must be selected such that $|\sigma_{XY}(\omega)| \le 1$ for all $\omega$.

Finally, to yield the third case where the bivariate coherence decays polynomially we propose
\begin{align}
\sigma_{XY}(\omega)&=\frac{\sigma_0}{\left[\left(\omega-\omega_0\right)^2+b^2\right]^{\alpha/2}},\;0<\sigma_0<
\left(\omega_0^2+b^2\right)^{\alpha/2}.
\end{align}
This form allows the smoothness of the cross-spectrum to be directly manipulated by varying $\alpha$, and is related to spatial modeling using the Mat\'ern covariance kernel~\cite{gneiting2010matern}.

Hermitian symmetry dictates that the group delay $\theta_{XY}(\omega)$ is required to be an odd function, and we propose the parametric form
\begin{equation}
\label{eq:theta}
\theta_{XY}(\omega)=
\theta_1 \omega+\theta_2 \omega^3+\dots +\theta_l \omega^{2l-1}.
\end{equation}
Note that while a constant phase is not valid across all frequencies, we may have a phase that is locally constant within particular frequency bands (see~\cite[p.137]{hamon1974spectral}), and where the polynomial form of~\eqref{eq:theta} may simply be considered a non-parametric approximation to the group delay. A function that depends on higher terms than linear are exhibiting a non-linear frequency-dependent time delay, a characteristic of a wave passing through a dispersive medium~\cite{hamon1974spectral}.

Next we ask what constraint is imposed on the bivariate coherence and group delay if we require the signal to be proper, such that $R_{ZZ}(\omega)=0$ in~\eqref{eq:SC}.  It follows from Table~\ref{Table:RealComplex} that we require $S_{XX}(\omega)=S_{YY}(\omega)$ and $\Re\{S_{XY}(\omega)\}=0$ for the signal to be proper. For the second condition, by inspecting~\eqref{eq:cart_coherency} and \eqref{amp-phase}, we see that we require that {\em at each} positive frequency $|\omega|$ where $S_{XX}(|\omega|)=S_{YY}(|\omega|)>0$, either that $\sigma_{XY}(|\omega|)=0$ {\em or} $\theta_{XY}(|\omega|)= \pm \pi/2$ and $\theta_{XY}(-|\omega|)= \mp \pi/2$, such that $X$ and $Y$ are strictly ``out of phase." The latter possibility follows from the fact that a zero-mean Gaussian signal is proper if and only if it is \textit{circular} (i.e. a signal whose properties are invariant under rotation), see~\cite{schreier2010statistical}. An example of such would be a deterministic signal which follows an exact circle, which corresponds to $\sigma_{XY}(\omega)=1$ and $\theta_{XY}(\omega)=\pm\mathrm{sgn}(\omega)\pi/2$. 
%%%%%%%%%%%%%%%%%%%%%%%%%%%%%%%%%%%%%%%%%%%%%%%%%%%
\subsection{Rotary coherency} \label{prelim-rotary}
We can alternatively specify a stochastic model for the coherency between rotary components $Z^+(t)$ and $Z^-(t)$. As we have chosen the two processes to be analytic, and thus supported only on positive frequencies, we only need to model the covariance between $d\Psi_{Z^+}(\omega)$ and $d\Psi_{Z^{-}}(\omega)$ for $\omega>0$. This is equivalent to the covariance between the increment processes of $d\Psi_{Z}(\omega)$ and $d\Psi^\ast_{Z}(-\omega)$, as can be seen from contrasting~\eqref{eq:CramerC} with \eqref{eq:CramerPM}. Therefore in addition to specifying the variance of $d\Psi_{Z^+}(\omega)$ and $d\Psi_{Z^-}(\omega)$, corresponding to specifying the rotary spectra $S_{++}(\omega)$ and $S_{--}(\omega)$, we only need to model their covariance at the same frequency, which corresponds to the rotary cross-spectrum given by
\begin{align}
S_{+-}(\omega)&=\rho_{\pm}(\omega)\left[S_{++}(\omega)S_{--}(\omega)\right]^{1/2},\nonumber\\
\rho_{\pm}(\omega)&=\sigma_{\pm}(\omega)e^{-\ri\theta_{\pm}(\omega)},
\label{eq:rotary_coherency}
\end{align}
where $|\rho_{\pm}(\omega)|\leq1$ is the {\em rotary coherency} of $Z^+$ and $Z^-$, $\sigma_{\pm}(\omega)$ is the rotary coherence, and $\theta_{\pm}(\omega)$ is the rotary group delay. 

As in the bivariate case, one may construct models for the rotary coherence $\sigma_\pm(\omega)$ and rotary group delay $\theta_\pm(\omega)$ analogous to~\eqref{eq:sigma0}--\eqref{eq:theta}, with the difference that these functions now only need to be supported over positive frequencies. This conveniently means that we are not constrained to require $\sigma_\pm(\omega)$ and $\theta_\pm(\omega)$ to be even and odd functions respectively. For the proposed model for $Z^+(t)$ and $Z^-(t)$ to be a valid Gaussian process, we now only require the three conditions:
\begin{enumerate}
\item $S_{++}(\omega)\ge0$ and $S_{--}(\omega)\ge0$ for all $\omega$,
\item $|\sigma_\pm(\omega)|\le1$ for all $\omega$, such that 1) and 2) together ensure the determinant of the spectral matrix $\bm{S}_\pm(\omega)$ in~\eqref{eq:SPM} is nonnegative for all $\omega$ (such that the spectral matrix itself is nonnegative definite), and
\item the spectral matrix $\bm{S}_\pm(\omega)$ is integrable.
\end{enumerate}

By inspecting Table~\ref{Table:RealComplex}, we observe a useful property. If $X(t)$ and $Y(t)$ are known to be uncorrelated with each other, we see that $S_{+-}(\omega)$ is real-valued for all $\omega$. This means that in such cases we only need to specify a real-valued model for $\rho_{\pm}(\omega)$. In such instances, the simplest positive and compactly supported frequency-dependent model for the rotary coherency (related to the model of~\eqref{eq:sigma0}) is
\begin{equation}
\rho_\pm(\omega)=
\begin{cases}
\left|a_0-a_1 \omega \right|, & 0<\omega \leq \omega_0,\quad \frac{a_0}{a_1}=\omega_0,\\
0, &\omega_0<\omega.
\end{cases}
\label{rhon}
\end{equation}
We use this simple rotary coherency model in our oceanographic example in Section~\ref{S:Simulations}. If more complex forms are required for the rotary coherency, we can replace \eqref{eq:sigma0} with an expansion using both even and odd terms. Similarly \eqref{eq:theta} can also be replaced by an expansion using both even and odd terms, already advocated as an approximation by~\cite{hamon1974spectral}.

The magnitude of the coherency specified by~\eqref{rhon} will have implications for 
representing the complex-valued signal as an ellipse~\cite{schreier2008polarization,walden2013rotary} and therefore has a geometric interpretation. The ellipse defined between the axis of $dZ(\omega)$ and $d Z^\ast(-\omega)$ will represent a line
if $\rho_\pm(\omega)=1$, and a circle if $\rho_\pm(\omega)=0$ (yielding propriety of $Z_t$). The intrinsic geometry of the complex-valued signal (e.g. the ellipse, circle or rectilinear motion mapped out by the trajectory) can therefore be seen as specifying $\sigma_\pm(\omega)$, while
temporal shifts and misalignments are encoded by $\theta_{\pm}(\omega).$

By inspecting Table~\ref{Table:RealComplex} we see that a proper complex-valued signal has no rotary coherency, i.e. $S_{+-}(\omega)=0$, which is a useful definition of propriety. This can be exploited to formulate a test for impropriety, as is detailed later in Section~\ref{impropriety}.  
Furthermore, for spatially driven processes such as oceanographic signals obtained from drifters, propriety 
is found to be related to the condition of spatial isotropy, with impropriety in turn implying anisotropy. Thus by testing for impropriety in the time series signal, we can test for anisotropy in the spatial process that generates the sampled signal. 

Modeling in bivariate or rotary components is of course equivalent, but as can be seen from Table~\ref{Table:RealComplex}, a parametric model represented in one decomposition will not in general be interpretable and compactly represented in the other---hence the need to specify classes of models in both decompositions.
%%%%%%%%%%%%%%%%%%%%%%%%%%%%%%%%%%%%%%%%%%%%%%%%%%%
%%%%%%%%%%%%%%%%%%%%%%%%%%%%%%%%%%%%%%%%%%%%%%%%%%%
\section{Parameter Estimation}\label{S:Whittle}
In this section we turn to the problem of estimating the parameters of a chosen model from observations. Maximum likelihood is a standard approach but is computationally slow requiring calculation of the determinant and inverse of the time-domain covariance matrix. Instead, \cite{whittle1953analysis} proposed using what later became known as the Whittle likelihood, which approximates the time-domain log-likelihood in the frequency domain in $\mathcal{O}(N\log N)$ operations, where $N$ is the length of the observed signal. In this section, we derive the correct form of the Whittle likelihood in each of the bivariate/complex/rotary representations, such that it can be easily implemented for any of the model specifications discussed in Section~\ref{S:Models}.
\subsection{Maximum Likelihood for Bivariate Processes}
Henceforth we let $X$ and $Y$ denote length~$N$ samples of the corresponding stochastic processes arranged as column vectors. To find the log-likelihood appropriate for the bivariate process consisting of $X$ and $Y$ we concatenate the two time domain samples into a single vector:
${B}^T=[{X}^T \; {Y}^T]$.
The theoretical $2N\times 2N$ covariance matrix for this vector under the assumed model is denoted by $\bm{C}_B(\bm{\theta})=\E\left\{BB^T\right\}$. The log-likelihood of the vector $B$ can then be written as
\begin{equation}
\label{log-likelihood2}
\ell(\bm{\theta})\overset{C}{=}-\frac{1}{2}\log|\bm{C}_B(\bm{\theta})|-\frac{1}{2}B^T\bm{C}_B^{-1}(\bm{\theta})B.
\end{equation}
The superscript ``$-1$" is the matrix inverse, $|\bm{X}|$ denotes the determinant of matrix $\bm{X}$, and $\overset{C}{=}$ denotes equality up to an additive constant which can be ignored as we are optimizing the objective function. The best choice of parameter-vector $\bm\theta$ for our chosen model to characterize the observed data is found by maximizing the log-likelihood function
\[
\widehat{\bm{\theta}}=\arg \max_{\bm{\theta}\in \bm{\Theta}} \ell(\bm{\theta}),
\]
where $\bm\Theta$ denotes the parameter space of $\bm\theta$. Maximum likelihood estimation (for Gaussian processes) is generally asymptotically efficient, and a well-behaved procedure in the time series setting \cite[p.27]{dzhaparidze1983spectrum}.

\subsection{The Whittle Likelihood for Bivariate Processes}
Computation of the matrix inverse and determinant in~\eqref{log-likelihood2} is computationally expensive, with complexity typically scaling as $\mathcal{O}(N^2)$ for stationary regularly-sampled processes (and $\mathcal{O}(N^3)$ more generally), where $N$ is the length of the signal. A standard technique to approximating~\eqref{log-likelihood2} is the Whittle likelihood \cite{whittle1953estimation}. This estimation technique approximates the time-domain log-likelihood function in the frequency domain, which results in improved $\mathcal{O}(N \log N)$ computational complexity. For bivariate signals, we first define the bivariate Discrete Fourier Transform (DFT) vector as
\[
J_B(\omega)=\begin{bmatrix}J_X(\omega)\\J_Y(\omega) \end{bmatrix}=\sqrt{\frac{\Delta}{N}}\sum_{t=1}^{N}\begin{bmatrix} X_t \\ Y_t \end{bmatrix}e^{-\ri\omega t\Delta},
\]
where $\Delta>0$ denotes the length of the sampling interval. The standard bivariate Whittle likelihood, once discretized, is given in~\cite{whittle1953analysis} by
\begin{multline}
\label{whittle_lik}
\ell_W(\bm{\theta})\overset{C}=\\-\frac{1}{2}\sum_{\omega\in \Omega}  \left[\log \left|\bm{S}_B(\omega;\bm{\theta}) \right| +J_B^\H(\omega)\bm{S}_B^{-1}(\omega;\bm{\theta})J_B(\omega)\right],
\end{multline}
where the subscript ``W" stands for ``Whittle," $\bm{S}_B(\omega)$ is as defined in~\eqref{eq:SB}, and $\Omega$ is the set of discrete Fourier frequencies: $\frac{2\pi}{N\Delta} (-\lceil N/2 \rceil +1,\ldots,-1,0,1,\ldots, \lfloor N/2 \rfloor)$. 
The optimal parameter choice for the Whittle likelihood is made by maximizing
\[
\widehat{\bm{\theta}}^{(W)}=\arg \max_{\bm{\theta}\in \bm{\Theta}} \ell_W(\bm{\theta}).
\]
This procedure is $O(N \log N)$ because $J_B(\omega)$ can be computed using a Fast Fourier Transform (FFT), the summation in~\eqref{whittle_lik} is over $\mathcal{O}(N)$ frequencies, and the matrix inverse of $S_B(\omega;\bm{\theta})$ now involves a $2\times2$ matrix, instead of the $2N\times 2N$ covariance matrix in~\eqref{log-likelihood2}.

It is known that the log-likelihood function $\ell_W(\bm{\theta})$ given in~\eqref{whittle_lik} approximates the bivariate log-likelihood function $\ell(\bm{\theta})$ given in~\eqref{log-likelihood2}, but in general $\widehat{\bm{\theta}}^{(W)}\neq\widehat{\bm{\theta}}$. The quality of the approximation (see~\cite{dzhaparidze1983spectrum}) depends on properties of the spectrum (as is clear from the results of~\cite{dahlhaus1988small}). The difference 
between the two objective functions specified by~\eqref{log-likelihood2} and~\eqref{whittle_lik} can be bounded, both in terms of mean and variance~\cite[Thm 5.2]{dzhaparidze1983spectrum}, and the rates achieved asymptotically in estimation using the Whittle likelihood is established by~\cite[Thm 5.5]{dzhaparidze1983spectrum}. Asymptotically, the same rates of convergence ($\sqrt{N}$) are achieved for the Whittle and standard log-likelihood. 

\subsection{The Whittle Likelihood for Complex-Valued Processes}\label{ss:whittlecomplex}
The Whittle likelihood for complex-valued processes has an identical form to the Whittle likelihood for bivariate processes, as is now shown. To find its form, we express~\eqref{whittle_lik} in terms of $Z\equiv X + \ri Y$. We start by defining the DFT for complex-valued processes given by
\[
J_C(\omega)=\begin{bmatrix}J_Z(\omega)\\J_{Z^\ast}(\omega) \end{bmatrix}=\sqrt{\frac{\Delta}{N}}\sum_{t=1}^{N}\begin{bmatrix}
Z_t \\
Z^\ast_t
\end{bmatrix}e^{-\ri\omega t\Delta}.
\]
There is a simple linear relationship between the DFTs $J_{C}(\omega)$
and $J_B(\omega)$
\begin{equation}
J_C(\omega)
=
\sqrt{2}\bm{T}J_B(\omega),
\label{eq:Jpm2Ju}\end{equation}
where $\bm{T}$ is the unitary matrix defined in~\eqref{eq:unitary}.
We then substitute~\eqref{eq:Jpm2Ju} and~\eqref{eq:B2C} into~\eqref{whittle_lik} such that
\begin{multline}
{\ell}_W(\bm{\theta})\overset{C}=-\sum_{\omega\in \Omega}\left[\log \left|\frac{1}{2}\bm{T}^\H\bm{S}_C(\omega;\bm{\theta})\bm{T}\right| \right.\\ \left.+J_C^\H(\omega)
\frac{\bm{T}}{\sqrt{2}} \left\{\frac{1}{2}\bm{T}^\H\bm{S}_C(\omega;\bm{\theta})\bm{T}\right\}^{-1}\frac{\bm{T}^\H}{\sqrt{2}}
J_C(\omega)\right]
\\ \overset{C}=-\frac{1}{2}\sum_{\omega\in\Omega}  \left[\log \left|\bm{S}_C(\omega;\bm{\theta})
\right| +J_C^\H(\omega)\bm{S}_C^{-1}(\omega;\bm{\theta})J_C(\omega)\right],
\label{whittle_complex}
\end{multline}
and this objective function is thus identical in appearance to the one provided in~\eqref{whittle_lik}. This result holds because $[Z \; Z^\ast]^T$ is a scaled unitary transformation of $[X\;Y]^T$, and would still hold if any scaled unitary matrix were substituted for $\bm{T}$.

We note that for proper signals, when $R_{ZZ}(\omega)=0$, this log-likelihood function takes the simplified form
\begin{equation}
\ell_W(\bm\theta)\overset{C}=-\sum_{\omega\in\Omega}\left[ \log\{S_{ZZ}(\omega;\bm\theta)\}+\frac{\hat{S}_{ZZ}(\omega)}{S_{ZZ}(\omega;\bm\theta)} \right],
\label{whittle_complex2}
\end{equation}
where $\hat{S}_{ZZ}(\omega)=|J_Z(\omega)|^2$ is the periodogram of the vector $Z$. This form occurs because the off-diagonal elements of $\bm{S}_C$ vanish, while the two main diagonal terms contribute identically, thus removing the factor of 1/2 from~\eqref{whittle_complex}. The log-likelihood function~\eqref{whittle_complex2} has a recognizable form which is similar to the Whittle likelihood for real-valued signals. In general however, the complementary spectrum must be considered and the form of~\eqref{whittle_complex} must be adopted.

\subsection{The Whittle Likelihood for Rotary Components}
\label{ss:whittlerotary}
We find the Whittle likelihood for rotary components $Z^+$ and $Z^-$ takes a slightly different form. We start by defining the rotary DFT vector by
\begin{equation}
J_\pm(\omega)=\begin{bmatrix}J_+(\omega)\\J_-(\omega) \end{bmatrix}=U(\omega)\begin{bmatrix}J_Z(\omega)\\J_Z(-\omega) \end{bmatrix},
\label{rotaryDFT}
\end{equation}
where the scalar function $U(\omega)$ is given in~\eqref{eq:unitstep}. It then follows from~\eqref{eq:C2PM} that~\eqref{whittle_complex} can be expressed in terms of rotary components by
\begin{equation}
\ell_W(\bm{\theta})\overset{C}=-\sum_{\omega\in\Omega^+}  \left[\log \left|\bm{S}_{\pm}(\omega;\bm{\theta})
\right| +J_{\pm}^\H(\omega)\bm{S}_{\pm}^{-1}(\omega;\bm{\theta})J_{\pm}(\omega)\right],
\label{whittle_rotary}
\end{equation}
where $\bm{S}_\pm(\omega;\bm\theta)$ is defined as in~\eqref{eq:SPM}. The summation only includes the frequencies $\omega\in\Omega^+$ where $\Omega^+$ is the set of positive discrete Fourier frequencies: $\frac{2\pi}{N\Delta} (1,\ldots, \lfloor N/2 \rfloor)$. This is because $\bm{S}_\pm$ vanishes for negative frequencies, which removes the need for the factor of 1/2 that accounts for the double-counting of contributions to the other two Whittle likelihood functions (\eqref{whittle_lik} and \eqref{whittle_complex}) at positive and negative frequencies. These multiplicative factors do not affect the optimization of the log-likelihood function, but are important when performing generalized likelihood ratio tests, as we shall demonstrate in Section~\ref{impropriety}. This form of the log-likelihood function allows the direct implementation and estimation of rotary coherence models specified in Section~\ref{prelim-rotary}, such as~\eqref{rhon}; and we therefore use~\eqref{whittle_rotary} in our practical example in Section~\ref{S:Simulations}. 

The Whittle likelihood for rotary components does not require the observed signal $Z$ to be explicitly split into rotary components; instead this split is performed implicitly from the definition of $J_\pm(\omega)$ in~\eqref{rotaryDFT}. If one does wish to obtain the analytic and conjugate analytic time series from $Z$ then this can be done through discrete Hilbert transforms (see also~\cite{marple1999computing})
\[
\begin{bmatrix} Z^+ \\ (Z^-)^\ast
\end{bmatrix}\equiv\frac{1}{\sqrt{2}}\bm{T}\begin{bmatrix} Z \\ \mathcal{H}Z
\end{bmatrix}=\frac{1}{2}\begin{bmatrix}(X - \mathcal{H}Y)+i(Y+\mathcal{H}X)\\(X + \mathcal{H}Y)+i(Y -\mathcal{H}X)\end{bmatrix},
\]
with $\mathcal{H}Z$ denoting the discrete Hilbert transform of $Z$, and similarly for $X$ and $Y$. The vectors $Z^+$ and $Z^-$ are not direct samples from the theoretical processes $Z^+(t)$ and $Z^-(t)$, but are instead discrete approximations to the continuous-time Hilbert transforms, rather like how the discrete Fourier transform is related to the Fourier transform.

Note that the frequency $\omega=0$ is not included in $\Omega^+$. This occurs because the rotary DFT~\eqref{rotaryDFT} is split evenly between the components at $\omega=0$, therefore the rotary coherency cannot be estimated at this frequency. This is unlikely to be a substantial drawback in practice, as the zero-frequency content of the signal is lost whenever the mean of the signal is removed prior to analysis, such that there is no contribution to the log-likelihood function at this frequency in any case. If it is important to include the zero-frequency component in the model, and the amplitude of the signal at this frequency is appreciably greater than zero, then the contribution to the log-likelihood function at this frequency can be added into~\eqref{whittle_rotary} by taking the zero frequency contribution from~\eqref{whittle_lik} or \eqref{whittle_complex}.

Finally,~\eqref{whittle_lik}, \eqref{whittle_complex}, or \eqref{whittle_rotary} can be combined with procedures to reduce known bias effects of the Whittle likelihood with moderate sample sizes~\cite{dahlhaus1988small}. These procedures include tapering the signal, for example with a cosine or Slepian taper~\cite{velasco2000whittle}; or by using the {\em de-biased} Whittle likelihood, proposed in~\cite{sykulski2016debiased} for real-valued signals. It has been observed that the bias reductions in these methods are often still significant for sample sizes of the order of 1,000 observed points, particularly with processes whose spectral densities are hard to approximate because of leakage and aliasing effects~\cite{percival1993spectral}.

%%%%%%%%%%%%%%%%%%%%%%%%%%%%%%%%%%%%%%%%%%%%%%%%%%%
\section{Statistical Procedures for Complex-Valued and Rotary Signals}\label{S:Implem}
In this section we discuss several useful practical procedures when stochastically modeling pairs of signals using the complex-valued or rotary representation.
%%%%%%%%%%%%%%%%%%%%%%%%%%%%%%%%%%%%%%%%%%%%%%%%%%%
\subsection{Semi-parametric modeling and estimation} \label{semipm}
Commonly signals are acquired to be both low-pass filtered and notch filtered~\cite{glover1977adaptive} before digitized, thus missing a range of frequencies. In other application areas, the proposed parametric model only fits over a range of frequencies~\cite{hamon1963estimating,hamon1974spectral,wahba1980automatic,robinson1994semiparametric,lumpkin2010surface,sykulski2016lagrangian}.
In these instances
\begin{equation}
S_{\pm}(\omega)=\begin{cases}
S_{\pm}(\omega;\bm{\theta}) & \omega \in \Omega_S\\
S_{\pm}(\omega) & \omega \notin \Omega_S,
\end{cases}
\end{equation}
where $\Omega_S$ is the range of frequencies in which $S_{\pm}(\omega)$ can be modeled parametrically from the data.

For all such instances it is not suitable to infer the parameters of the generating mechanism of the continuous time process using~\eqref{whittle_complex} or \eqref{whittle_rotary} when any of the Fourier frequencies $\frac{2\pi}{N\Delta} (1,\ldots, \lfloor N/2 \rfloor)\not\in \Omega_S$. The log-likelihood function for $\bm{\theta}$ in~\eqref{whittle_rotary} should instead be restricted to the Fourier frequencies $\omega\in \Omega_S$ such that
\begin{equation}
\ell_W(\bm{\theta})\overset{C}=-\sum_{\omega\in \Omega_S}  \left[\log \left|S_{\pm}(\omega;\bm{\theta})
\right| +J_{\pm}^\H(\omega)S_{\pm}^{-1}(\omega;\bm{\theta})J_{\pm}(\omega)\right],
\label{whittle_rotary2}
\end{equation}
and equivalently with~\eqref{whittle_complex}.
In fact, for the case of~\cite{hurvich1998mean} the optimal number of frequencies to use in estimation can be determined and is not $\sim N$ but rather scales as $\sim N^{4/5}$. In general we would expect the precision of parameter estimates to scale as $|\Omega_S|^{-1/2}$, where $|\Omega_S|$ denotes the number of Fourier frequencies contained in $\Omega_S$, thus loosing in precision for the reduced bias.
 
 This means that only the frequencies used in the log-likelihood summation determine the parameter estimates, and all other sampled frequencies are excluded. In such instances the resulting parameter estimation procedure is {\em semi-parametric}~\cite{robinson1995gaussian}. For set problems, such as that treated in~\cite{robinson1994semiparametric,hurvich1998mean}, the subset can be determined analytically prior to analysis. Sometimes high frequencies will be omitted to account for sampling errors such as measurement noise. This effect is well-documented in econometrics, finance and
geostatistics~\cite{olhede2009frequency,cressie1993statistics}, and while noise often
affects all frequencies equally, the signal is usually concentrated at low-frequencies. Additionally, effects due to sampling are sometimes stronger in their contamination of higher frequencies~\cite{masry1978alias}.

We can also use the semi-parametric approach to model only one of the rotary components, $Z^+(t)$ or $Z^-(t)$, akin to modeling only one side of the spectrum in the complex representation. This can be performed, for instance, when the physical process of interest is known to only spin in one direction for a given signal. This can simplify
analysis when one side of the spectrum is known to be contaminated by 
nuisance effects~\cite{sykulski2016lagrangian}, and while excluding these frequencies will increase variance, the bias will be decreased by a more significant amount, thus reducing overall estimation error.
%%%%%%%%%%%%%%%%%%%%%%%%%%%%%%%%%%%%%%%%%%%%%%%%%%%
\subsection{Hypothesis testing for impropriety}\label{impropriety}
In Section~\ref{prelim-rotary}  we specified models for the coherency between the rotary components $Z^+(t)$ and $Z^-(t)$. We may wish to test a proposed model for coherency against the simpler scenario where $\rho_\pm(\omega)=0$ for all $\omega$ in~\eqref{eq:rotary_coherency}. This hypothesis corresponds to there being no relationship between the positive rotating and negative rotating phasors. We note from Table~\ref{Table:RealComplex} that this hypothesis is {\em not} equivalent to $S_{XY}(\omega)=0$; rather, a process with zero rotary cross-spectra, $S_{+-}(\omega)=0$, is equivalent to a proper process, $R_{ZZ}(\omega)=0$ for all $\omega\neq0$. This is a deliberate and convenient consequence of the way we have constructed $Z^+(t)$ and $Z^-(t)$ to {\em both} be analytic. Testing for impropriety in the time domain is then equivalent to testing for a non-zero cross-spectrum between rotary components in the frequency domain.

There are many tests for impropriety using data sets of multiple replicates of complex-valued vectors, see for example \cite{schreier2010statistical,ollila2004generalized,walden2009testing,chandna2016frequency,tugnait2016testing}. We note that these are {\em non-parametric tests} in general requiring {\em replicated} complex-valued vectors. Here, by contrast, we have a simple {\em parametric} time series model suitable for a {\em single time series}, and will derive a test statistic for this scenario. 

Given a chosen model for $S_{++}(\omega)$ and $S_{--}(\omega)$, we
estimate the coherency structure by specifying a parametric model for $\rho_{\pm}(\omega)$ in \eqref{eq:rotary_coherency}. We may however prefer to use the simpler model where $\rho_{\pm}(\omega)=0$, such that $S_{+-}(\omega)=0$. We can make use of the parametric model---in combination with the Whittle likelihood for complex-valued signals---to perform a generalized likelihood ratio test to check for evidence if $\rho_{\pm}(\omega) \neq 0$ such that $S_{+-}(\omega)\neq0$. Specifically, this is done by computing the generalized likelihood ratio test statistic
\begin{equation}
W = 2\left[\ell_W(\widehat{\bm\theta}_1)-\ell_W(\widehat{\bm\theta}_0)\right],
\label{e:LRstatistic}
\end{equation}
where $\widehat{\bm\theta}_1$ includes non-zero estimates for the coherency $\rho_{\pm}(\omega)$, while for $\widehat{\bm\theta}_0$, $\widehat\rho_{+-}(\omega)=0$ for all $\omega$.
The estimators $\widehat{\bm\theta}_1$ and $\widehat{\bm\theta}_0$ are obtained by maximizing~\eqref{whittle_rotary}. We assume that the full model with non-zero coherence has $p$ extra parameters that are not linearly dependent. For example,
in the case of the coherency structure proposed in~\eqref{eq:sigma} and \eqref{eq:theta}, then $p=r+l+1$. Whereas in~\eqref{rhon} if we set $a_1=0$ for example, such that we have
a fixed constant for $\rho_{\pm}(\omega)$ when $\omega<\omega_0$, then we have $p=1$.

In more generality, if we use the time domain bivariate log-likelihood function of~\eqref{log-likelihood2} in~\eqref{e:LRstatistic}, then from standard likelihood theory, the test statistic $W$ is asymptotically distributed according to a $\chi^2_p$ distribution with degrees of freedom equal to the number of extra parameters $p$ in the alternative hypothesis versus the null. In Appendix~\ref{SM:HT}, we prove that the Whittle likelihood for complex signals also asymptotically yields a test statistic with a $\chi^2_p$ distribution under the null, for special cases where the rotary spectra are equal in magnitude across frequency, but possibly correlated with each other such that the process is improper.
%%%%%%%%%%%%%%%%%%%%%%%%%%%%%%%%%%%%%%%%%%%%%%%%%%%\
\subsection{Model choice}\label{modelchoice}
In many practical scenarios the most appropriate choice for a model, and the corresponding number of parameters, is unknown \textit{a priori}. When it is possible to define a model such that possible candidate models are {\em nested}---that is, with simpler models being recovered from a more complete model by setting certain parameters to constants---then it is possible to apply the method of {\em model choice}~\cite{vuong1989likelihood}. In this section, we provide the correct form of various model choice procedures for complex-valued signals.

The Akaike Information Criterion (AIC) is a model choice procedure that can be used, for example, to find the most appropriate choice of $q_1$ and $q_2$ for a range of ARMA$(q_1,q_2)$ processes~\cite{hurvich1989regression}. Alternatively, the AIC can be used to select the order of the coherency models proposed in~\eqref{eq:sigma0} or \eqref{eq:sigma}. For complex-valued signals, the AIC can be closely approximated in the frequency domain by using the Whittle likelihood to approximate the time-domain log-likelihood, and combining this with a model penalization parameter as in \cite{hurvich1989regression} it follows that it takes the same form as for real-valued signals
\[
\textrm{AIC}(\bm{\hat\theta})=-2\ell(\bm{\hat\theta})+2q\approx-2\ell_W(\bm{\hat\theta}^{(W)})+2q,
\]
where $q$ is the number of parameters in the model. The model with the smallest AIC value is then selected. 

For small sample sizes, it is often recommended to use a correction to the AIC known as the AICC \cite{hurvich1989regression}, which for complex-valued signals is given by
\begin{align}
\textrm{AICC}(\bm{\hat\theta})&=-2\ell(\bm{\hat\theta})+\frac{4qN}{2N-q-1}\nonumber\\&\approx-2\ell_W(\bm{\hat\theta}^{(W)})+\frac{4qN}{2N-q-1},
\label{eq:AICC}
\end{align}
which converges to the AIC for large sample sizes. The correction uses $2N$ (rather than $N$ as would be done for real-valued signals), as there are $2N$ degrees of freedom in a length $N$ complex-valued signal. In the case of semi-parametric estimation as discussed in Section~\ref{semipm},~\eqref{eq:AICC} should not be used in this exact form, as the sample-size correction needs to take account of degrees of freedom in the data not used in the estimation. We instead replace $N$ by $|\Omega_S|$ in~\eqref{eq:AICC}, which is the number of Fourier frequencies in $\Omega_S$ in~\eqref{whittle_rotary2}.

When a tapered spectral estimate is used in the Whittle likelihood, then the degrees of freedom are further reduced by the correlation induced in neighboring Fourier frequencies by the taper~\cite{percival1993spectral}. This loss of resolution is also known as narrowband blurring, in contrast to the broadband blurring which is attributed to the leakage of the Fej\'er kernel in the periodogram for example. Without tapering, nominally each pair of Fourier coefficients, spaced $1/(N\Delta)$ apart, are uncorrelated; however, with tapering, coefficients spaced $\gamma/(N\Delta)$ apart are approximately uncorrelated, where $\gamma>1$ is some constant reflecting the implicit bandwidth of frequency-domain smoothing. We must then further reduce $N$ (or $|\Omega_S|$ for semi-parametric estimation) by a factor of $\gamma$ in the AICC, which can be computed by Fourier transforming the taper onto a fine grid and then finding the frequency width where the transformed sequence is only correlated below some specified threshold.
%%%%%%%%%%%%%%%%%%%%%%%%%%%%%%%%%%%%%%%%%%%%%%%%%%%%%%%%%%%%%%%%%%%%%%%%%%%%%%%%%%%%%%%%%%%%%%%%%%%%%%
\section{Application to turbulent flow data}\label{S:Simulations}
In this section we present an application of the modeling and estimation methods we have developed for bivariate and complex-valued signals. All results can be exactly reproduced using MATLAB code freely downloadable from \texttt{http://} \texttt{ucl.ac.uk/statistics/research/spg/software}, which also includes a technical description of how the turbulent flow data is numerically generated. The numerical simulation can be reproduced using software available at \texttt{http://jeffreyearly.com/numerical-models/}

\begin{figure*}
\centering{
\includegraphics[width=0.8\textwidth]{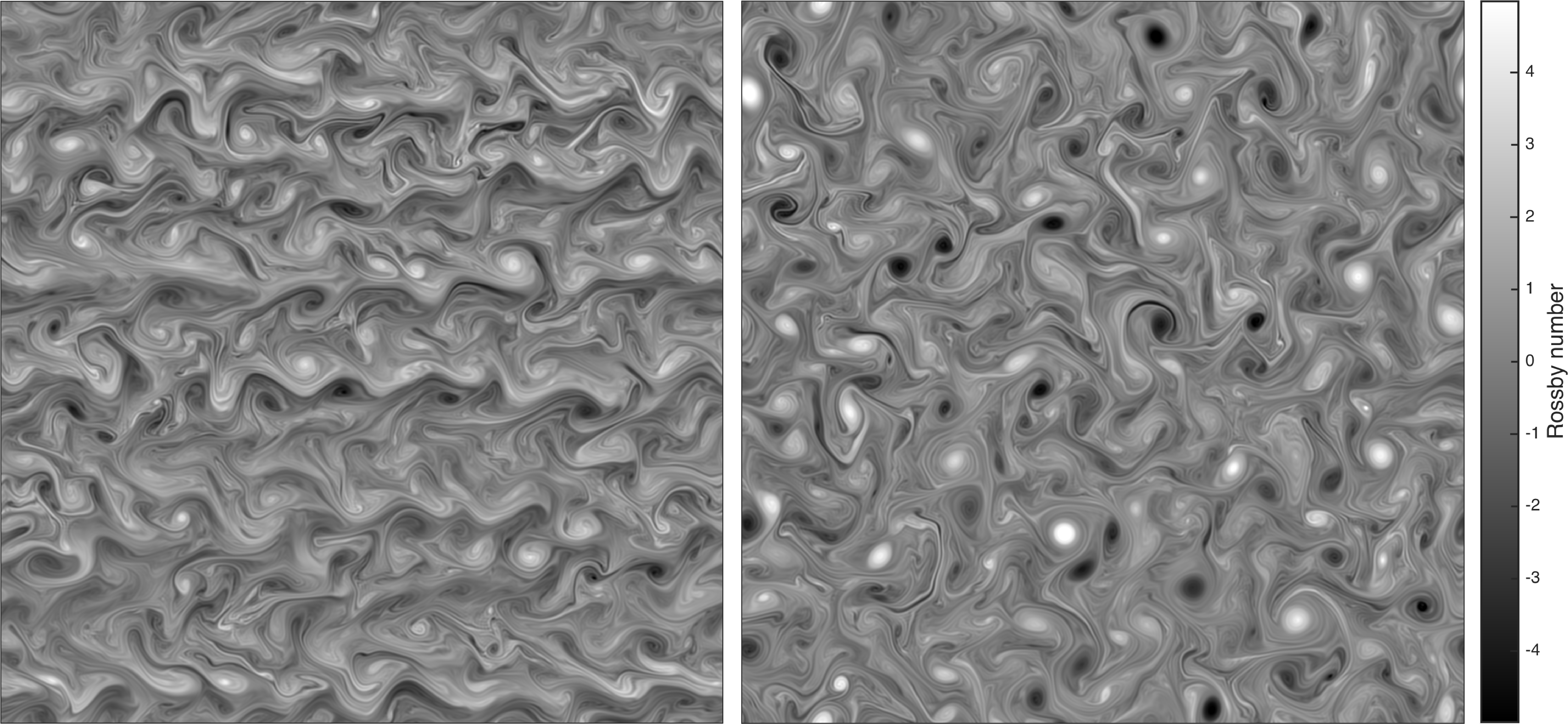}}
\caption{\label{FigVorticities}Snapshots from numerical experiments for anisotropic (left) and isotropic (right) two-dimensional turbulence. The field presented is relative vorticity, the curl of the velocity vector field at each point.  East-west bands are apparent in the anisotropic simulation that do not appear in the isotropic simulation. The color scale shows the Rossby number, a non-dimensional measure of the vorticity strength.}
\end{figure*}

We test our modeling and estimation procedures on data obtained from quasi-geostrophic turbulent flow simulations. Specifically we track the spatial trajectories of 256 particles over time from two sets of numerical experiments, snapshots of which are displayed in Fig.~\ref{FigVorticities}. Here trajectories of particles are computed from the time-varying velocity fields associated with quasi-steady state forced-dissipative two-dimensional turbulence simulations under both isotropic ($f$-plane) and anisotropic ($\beta$-plane) dynamics. Such simulations are standard in oceanography, see e.g. \cite{vallis2006atmospheric} for details.

We represent the velocities of the Lagrangian trajectories as complex-valued signals, which we denote as $z=u+\ri v$. The time between observations of the signal is set to 1 day. We model $z$ as following a Mat\'ern process, as motivated in \cite{lilly2016fractional}, which in the isotropic case has rotary spectra defined by
\begin{equation}
S_{++}(\omega) = S_{--}(\omega) = \frac{\phi^2}{(\omega^2+\alpha^2)^{\nu+1/2}}, \quad \omega>0,
\label{eq:Isotropy}
\end{equation}
where $\omega$ is given in cycles per day. The smoothness parameter $\nu>0$ defines the Hausdorff dimension of the graph---equal to max$(1,2-\nu)$---as well as the degree of differentiability of the process. The range parameter $\alpha>0$ is a timescale parameter, where $1/\alpha$ can be referred to as the correlation timescale, and $\phi^2>0$ defines the magnitude of the variability of the process. The motivation for using the Mat\'ern for turbulence velocities is based on its properties: both the Mat\'ern and Lagrangian trajectories such as these exhibit close to power law behavior at mid-to-high frequencies (see also the review paper of \cite{lacasce2008statistics}). Moreover, in contrast to fractional Brownian Motion, the power law behavior of a Mat\'ern breaks at low frequency making the process stationary (and not self-similar)---this is again consistent with the discussions of \cite{lilly2016fractional,lacasce2008statistics}. We choose to model the two rotary components as Mat\'ern processes with the same parameters, as in this experiment there is no preferred or dominant direction of rotation of the particles, such that the rotary spectra are expected to be symmetric.

\begin{figure}
\centering
\includegraphics[width=0.38\textwidth]{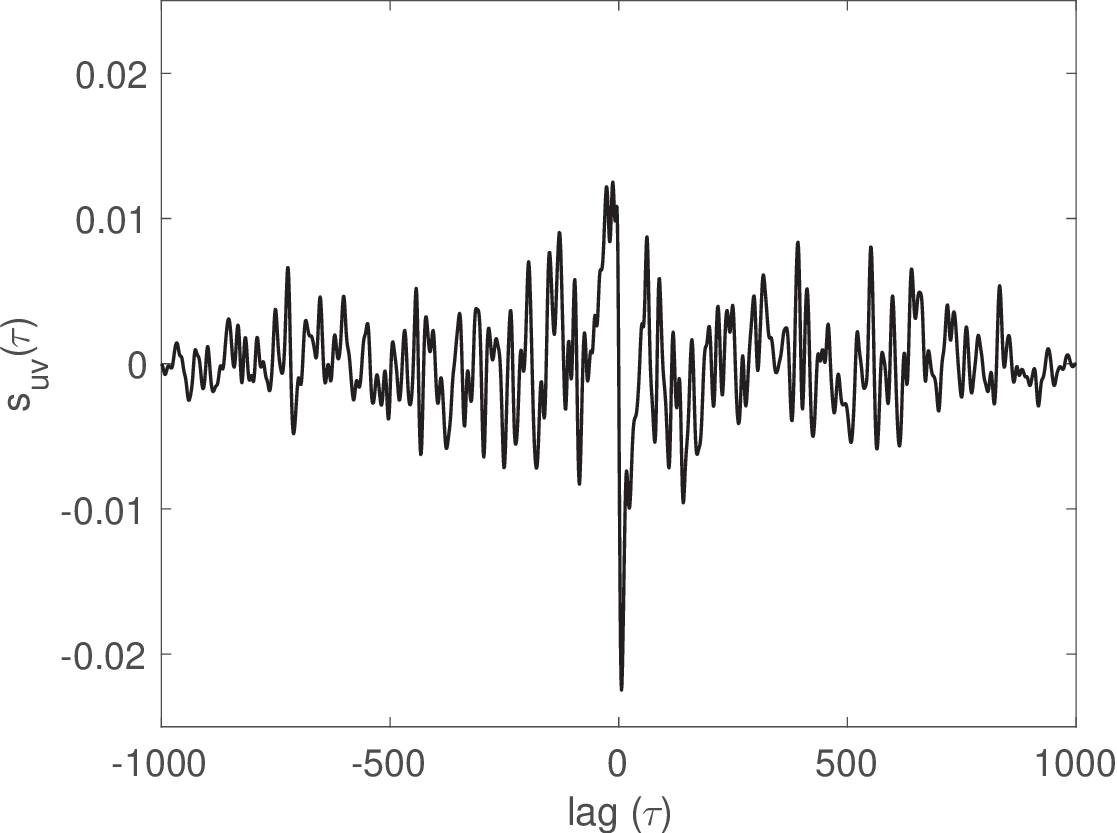}
\caption{\label{FigCov}The average autocorrelation of $u$ and $v$ across all observable lags of the signals analyzed from the anisotropic experiment displayed in Fig.~\ref{FigVorticities} (left). The estimated autocorrelation sequence has been averaged across the 256 signals analyzed in Section~\ref{S:Simulations}. The autocorrelations are estimated using the biased autocovariance estimator to reduce variance, where each lag is normalized by $N$ rather than $N-\tau$.}
\end{figure}

We account for potential anisotropy by modeling the rotary coherency. We employ the model of~\eqref{rhon} setting $a_0=1$, such that the coherency approaches unity as $\omega$ approaches zero. The justification for this is physical, as at long timescales we expect the east-west bands in the left panel of Fig.~\ref{FigVorticities} to be entirely dominant, which is consistent with a fully improper/anisotropic model with rotary coherency equal to unity. This leads to the following 1-parameter model
\begin{align}
S_{+-}(\omega)&= \rho_{\pm}(\omega)\left[S_{++}(\omega)S_{--}(\omega)\right]^{1/2}
\nonumber\\  \rho_\pm(\omega)&=\textrm{max}(0,1-c\omega),\quad\omega\ge0,
\label{eq:Anisotropy}
\end{align}
where $c\in\mathbb{R}^+$. The choice of compactly supporting $\rho_\pm(\omega)$---as discussed in Section~\ref{prelim-rotary}---is also physical, where isotropic behavior is expected at high frequencies beyond some physical timescale. Finally, as also discussed in Section~\ref{prelim-rotary}, we ignore the group delay and model $\rho_\pm$ as real-valued, which is reasonable if the signal components, in this case $u$ and $v$, are uncorrelated with each other. To check this, in Fig.~\ref{FigCov} we display the normalized cross-covariance between $u$ and $v$ averaged across 256 trajectories (each of length 1,001) from the numerical model, where it can be seen that it is reasonable to make the assumption that $u$ and $v$ are uncorrelated, as the estimated correlation never exceeds 0.025 at any lag.

\begin{figure*}
\centering
\includegraphics[width=0.8\textwidth]{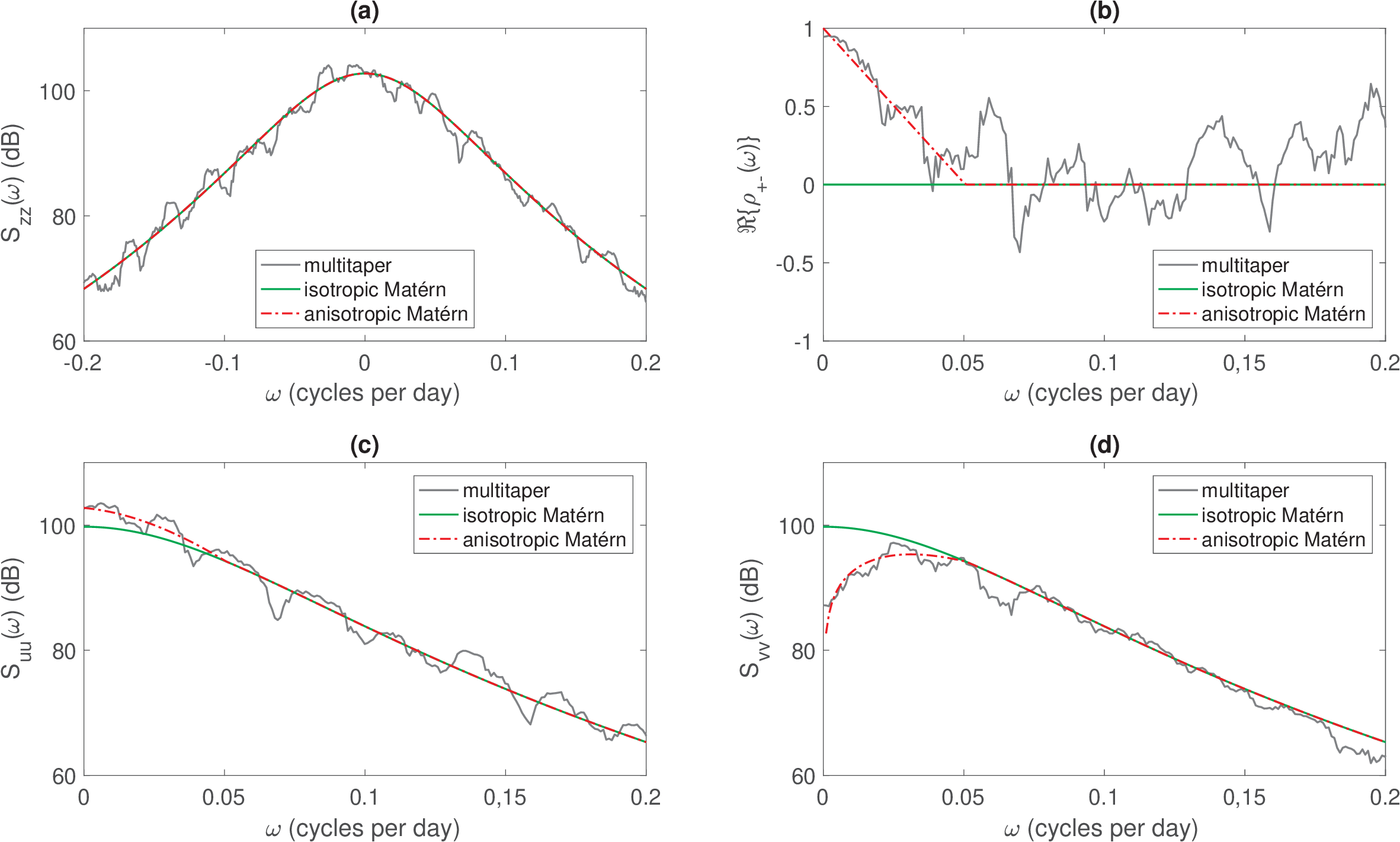}
\caption{\label{FigAnisoFit}
Spectra from the isotropic and anisotropic Mat\'ern model, with parameters estimated from the Whittle likelihood, plotted against non-parametric multi-taper spectral estimates from a complex-valued velocity signal observed from the anisotropic numerical simulation shown in the left panel of Fig.~\ref{FigVorticities}. Panel (a) is the Mat\'ern fit to the power spectrum $S_{ZZ}(\omega)$, which is identical for the isotropic and anisotropic models. Panel (b) displays the fit to the anisotropic model of $\rho_\pm(\omega)$ defined in \eqref{eq:Anisotropy}, where the isotropic model is zero at all frequencies by definition. Panels (c) and (d) display the anisotropic and isotropic model fits to the power spectrum of the velocities in the $u$ and $v$ direction only. The multi-tapers used are the discrete prolate spheroidal sequences (dpss) with bandwidth parameter set to 3.}
\end{figure*}

The full stochastic model therefore has four parameters: three for the Mat\'ern process, and one to specify the rotary coherency. The proposed model is a valid Gaussian process as $S_{++}(\omega)\ge0$, $S_{--}(\omega)\ge0$, $|\rho_\pm(\omega)|\le1$ for all $\omega>0$, and the spectral matrix $\bm{S}_\pm(\omega)$ is integrable, as discussed in Section~\ref{prelim-rotary}. The choice of the ``triangle function" for the coherency in~\eqref{eq:Anisotropy} naturally defines a timescale at which anisotropy begins, as the coherence is zero at $\omega = 1/c$, corresponding to a period of $2\pi c$, and then increases linearly as frequency decreases until at long timescales all energy is in the dominant Cartesian component, which is $u$. The model of~\eqref{eq:Anisotropy} defines the rotary coherency and its parameter can be estimated, together with the parameters of the Mat\'ern process~\eqref{eq:Isotropy}, using the Whittle likelihood for rotary components as provided in Section~\ref{ss:whittlerotary}. The parameters in the optimization are initialized using least squares, namely starting from~\eqref{eq:Isotropy} we first assume $\alpha=0$ and rewrite
\begin{equation}
\log S_{++}(\omega)=\log \phi^2-(2\nu+1)\log(\omega),
\end{equation}
and regress the observed log-periodogram $\log \hat{S}_{++}(\omega)$ over mid-range frequencies to make a least squares fit of $\nu$ and $\phi$. The parameter $\alpha$ is then set to a mid-range value of 0.1 cycles per day, and $c$ is set by finding the lowest frequency at which the estimated rotary coherency from multi-taper spectral estimates is zero. More details can be found in the online code.

In Fig.~\ref{FigAnisoFit}, we display the Whittle likelihood fit
to an individual signal (of length 1,001) obtained from the anisotropic numerical simulation shown in the left panel of Fig.~\ref{FigVorticities}. Because of the steep energy roll-off, we have used a semi-parametric fit by excluding 60\% of the frequency range, and thus only modeling up to 0.2 cycles per day, where the Nyquist is 0.5 cycles per day. We have also excluded the zero frequency from the fit, as we have removed the sample mean, and hence there is no spectral content at $\omega=0$. As the spectral slopes are steep, we use the tapered version of the Whittle likelihood~\cite{dahlhaus1988small}, as discussed in Section~\ref{S:Whittle}.
Specifically we estimate the parameters of the modeled spectra using non-parametric multi-taper spectral estimates obtained from discrete prolate spheroidal sequences (dpss) \cite{thomson1982spectrum}, otherwise known as Slepian tapers, with bandwidth parameter set to 3.

Analyzing the fits in Fig.~\ref{FigAnisoFit}, it can be seen that the extra parameter in our model has succinctly captured the difference between the flow in each Cartesian component ($u$ and $v$) at low frequencies. Thus the Mat\'ern model of~\eqref{eq:Isotropy}, plus rotary coherency as modeled in~\eqref{eq:Anisotropy}, appears to generally be a good fit for this complex-valued signal. We note that the estimated rotary coherency from multi-tapers is a noisy estimate, as compared with the estimate of the rotary power spectra contained in $S_{ZZ}(\omega)$. This is expected, particularly at higher frequencies, as the estimated coherency is measured from an individual signal, and at higher frequencies is often the ratio of two small but noisy quantities. Despite this, the parametric model for rotary coherency appears to have obtained a reasonable estimate, and captured the frequency-dependent structure of the coherency, particularly at low frequencies.

To put our choice of model under further scrutiny, we fit this 4-parameter anisotropic Mat\'ern model to all of the 256 trajectories from the anisotropic experiment and compare the value of the log-likelihood function versus the null hypothesis of a three-parameter isotropic Mat\'ern model, in which the rotary coherency is identically zero ($\rho_\pm(\omega)=0$). We can then use a generalized likelihood ratio test, as described in Section~\ref{impropriety}, to test for evidence of anisotropy. For consistency, we also repeat this procedure with the same number of trajectories (of the same length) from the isotropic experiment in the right panel of Fig.~\ref{FigVorticities}, to see if we correctly do not reject the null in such cases. While the null model is nested within the alternate model, the null value of the parameter $c$ is at the boundary of its range, requiring adjustments to calculate the critical value for the test statistic. Therefore, we compute 95\% confidence intervals for our test statistic by bootstrapping a number of isotropic simulated Mat\'erns (with parameters similar to those estimated in Fig.~\ref{FigAnisoFit}), rather than using a chi-squared statistic as derived in the Appendix.

\begin{figure}
\centering
\includegraphics[width=0.45\textwidth]{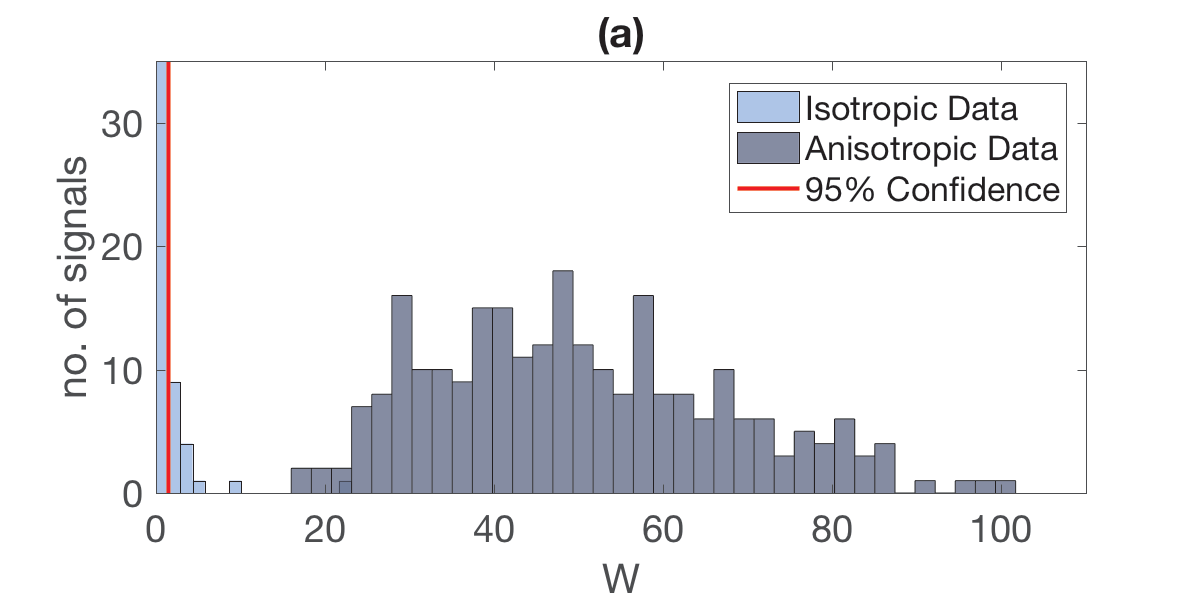}\vspace{3mm}
\includegraphics[width=0.45\textwidth]{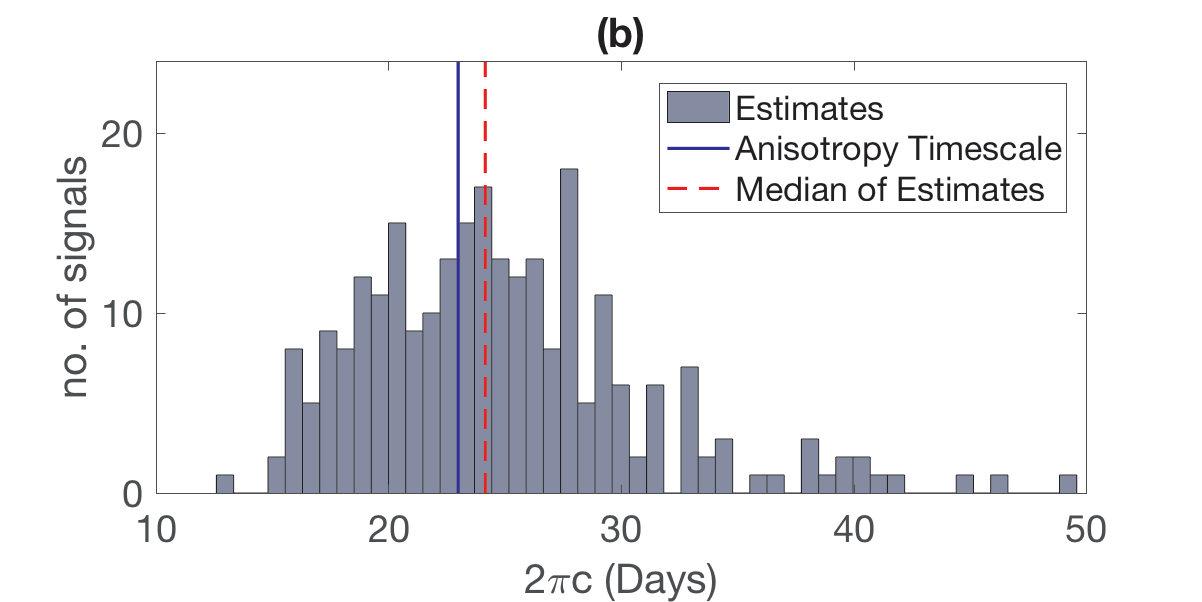}
\caption{\label{FigAnisoHist}(a) generalized likelihood ratio test statistics computed from 256 isotropic and anisotropic trajectories where the first column contains 240 values. The 95\% confidence interval is obtained from bootstrapping. (b) anisotropy timescale estimates $2\pi\hat{c}$ from the anisotropic trajectories.}
\end{figure}

The set of test statistics, calculated from~\eqref{e:LRstatistic}, from both experiments is displayed in Fig.~\ref{FigAnisoHist}(a). The statistics are compared with the 95\% one-sided bootstrapped confidence interval. The isotropic model is correctly always rejected for the anisotropic data and rejected only 16 times for the isotropic data (6.25\% of the 256 observed signals). This is in broad agreement with a type I error level set to 5\%. 
In experimental rather than controlled data we would account for multiple testing issues if performing this test over multiple trials, applying techniques such as False Discovery Rates (FDR).

The timescale associated with the observed anisotropy is of interest. As the signals were observed from a numerical model, we can assess which frequencies are associated with the anisotropic behavior from the settings. A spatial scale known as the Rhines scale \cite{rhines1975waves}, well-known in oceanography, determines the scale at which the transition to anisotropic large-scale behavior begins; this can be converted to a temporal scale through a division by the root-mean-square velocity.  This gives a time-scale of approximately 23 days for the anisotropic experiment. Fig.~\ref{FigAnisoHist}(b) provides estimates of this timescale for each signal from our parametric model, based on the estimate of the frequency at which the rotary coherence becomes zero, which is $2\pi c $. The median value is found to be 24.2 days, consistent with the 23 days computed from the Rhines scale. This apparent ability to infer a key {\em spatial} scale based solely on the {\em frequency} structure of time signals obtained Lagrangian trajectories is an interesting result showing the power of this method. Translating the temporal content garnered from multiple signals using our methods, to local spatial summaries, is an important avenue for future work.

%%%%%%%%%%%%%%%%%%%%%%%%%%%%%%%%%%%%%%%%%%%%%%%%%%%%%%%%%%%%%%%%%%%%%%%%%%%%%%%%%%%%%%%%%%%%%%%%%%%%%%%%%%%%%%%%%%%%%%%%%%%%%%%%%%%%%%%%%%%%%
\section{Conclusions}\label{S:Conclusions}
In this paper we have proposed a framework for stochastic modeling and estimation of stationary bivariate of complex-valued signals. We have shown the power of separating out behavior in frequency using the rotary components, and modeling different ranges of frequencies separately. This permitted us to handle a plethora of different effects directly in the frequency domain, and introduce new signal characteristics. For example, we demonstrated how our techniques can be used to effectively capture anisotropy in oceanographic flow models. In addition, we have proposed appropriate computationally-efficient parameter estimation procedures by extending the Whittle likelihood objective function to complex-valued and rotary signals.

There remain significant challenges in the frequency domain analysis of bivariate and complex-valued signals. A key challenge is to extend our modeling framework to nonstationary and higher order processes, where advances in non-parametric modeling have been made in \cite{schreier2010statistical}. The main application challenge is to continue building models from our framework for use in a wide range of physical applications, as we have performed recently in \cite{sykulski2015improper} with seismic data signals.
%%%%%%%%%%%%%%%%%%%%%%%%%%%%%%%%%%%%%%%%%%%%%%%%%%%
%%%%%%%%%%%%%%%%%%%%%%%%%%%%%%%%%%%%%%%%%%%%%%%%%%%%
\appendices
\section{Hypothesis testing for impropriety - proof of $\chi^2_p$-distributed test statistic}\label{SM:HT}
\subsection*{Hypothesis Test Set-Up}
We shall consider the special cases of 
\begin{align*}
{\mathbf{S}}_{\pm}^{(0)}\left(\omega;\bm{\theta}_0 \right)&=S\left(\omega;\bm{\theta}_M\right)
\cdot
\begin{bmatrix}
1 & 0\\
0 & 1
\end{bmatrix},\quad {\mathrm{vs}}\\
{\mathbf{S}}_{\pm}^{(1)}\left(\omega;\bm{\theta}_1\right)&=S\left(\omega;\bm{\theta}_M\right)
\cdot
\begin{bmatrix}
1 & \rho\left(\omega;\bm{\psi}\right)\\
\rho\left(\omega;\bm{\psi}\right) & 1
\end{bmatrix}.
\end{align*}
Where the null and alternate parameters are denoted as
\[
\bm{\theta}_0=\begin{bmatrix}
\bm{\theta}_M\\
\bm{\psi}_o
\end{bmatrix},\quad {\mathrm{vs}}\qquad \bm{\theta}_1=\begin{bmatrix}
\bm{\theta}_M\\
\bm{\psi}
\end{bmatrix}.
\]
We shall now develop a test for the null hypothesis of
\begin{align*}
&H_0:\quad {\mathbf{S}}_{\pm}\left(\omega\right)={\mathbf{S}}_{\pm}^{(0)}\left(\omega;\bm{\theta}_0 \right),\quad {\mathrm{vs}} \\ &H_1:\quad {\mathbf{S}}_{\pm}\left(\omega\right)={\mathbf{S}}_{\pm}^{(1)}\left(\omega;\bm{\theta}_1 \right).
\end{align*}
We note that the hypotheses are nested, and that makes standard theory possible---however this is not a necessary requirement~\cite{vuong1989likelihood}. Assume that $\bm{\theta}_M$ is an $l$-vector and $\bm{\psi}$ is a $p$-vector.

\subsection*{Notation for Quadratic Approximations}
Let us first define  the Fisher information matrix
\[
{\mathcal F}(\bm{\theta}_1)=\begin{bmatrix}
\tilde{\mathcal F}(\bm{\theta}_1) & {\mathcal F}_{\times}(\bm{\theta}_1)\\
{\mathcal F}_{\times}^T(\bm{\theta}_1)&{\mathcal F}_{\circ}(\bm{\theta}_1)
\end{bmatrix},
\]
where we have decomposed ${\mathcal F}(\bm{\theta}_1)$ into 4 blocks to simplify calculations: $\tilde{\mathcal F}(\bm{\theta}_1)$ contains the negative of the expected mixed second derivatives of the log-likelihood function with respect to the parameters in $\bm{\theta}_M$,  ${\mathcal F}_{\times}(\bm{\theta}_1)$ is the matrix with entries of the negative of the expectation of the second cross-derivatives between $\bm{\theta}_M$ and $\bm{\psi}$, and finally 
${\mathcal F}_{\circ}(\bm{\theta}_1)$ contains the negative of the expectation of the mixed second derivatives of the log-likelihood function with respect to the parameters in $\bm{\psi}$. 
In parallel we define the 
observed Fisher information matrices as ${\bm{F}}$, which has been decomposed into
$\tilde{\bm{F}}$, containing the negative of mixed second derivatives of the log-likelihood function with respect to the parameters in $\bm{\theta}_M$,  ${\bm{ F}}_{\times}(\bm{\theta}_1)$ is the matrix with entries of the negative of second cross-derivatives between $\bm{\theta}_M$ and $\bm{\psi}$, and finally 
${\bm{ F}}_{\circ}(\bm{\theta}_1)$ contains the negative of mixed second derivatives of the log-likelihood function with respect to the parameters in $\bm{\psi}$. Note that the observed Fisher information, and its expectation, as well as the corresponding submatrices are all a function of $\bm{\theta}_1=[ \bm{\theta}_M \,\, \bm{\psi}]^T$, where we recall also that $\bm{\theta}_0=[ \bm{\theta}_M \,\,\bm{\psi}_o]^T$.
We now also write 
\[
\hat{\bm{\theta}}_0=\begin{bmatrix}
\hat{\bm{\theta}}_{0M}&
\bm{\psi}_o
\end{bmatrix}^T,\quad {\mathrm{and}}\qquad \hat{\bm{\theta}}_1=\begin{bmatrix}
\hat{\bm{\theta}}_{1M} &
\hat{\bm{\psi}}
\end{bmatrix}^T,
\]
for the parameter values that maximize \eqref{whittle_rotary}, with the constraints of $H_0$, or without such constraints respectively. 

\subsection*{Properties of the Whittle Estimators}
We have already assumed that ${\bm F}\left( {\bm{\theta}}_0\right)$ is a continuous function of
${\bm{\theta}}_0$. We note that under the null $\hat{\bm{\theta}}_0\overset{P}{\rightarrow}
{\bm{\theta}}_0,$ and $\hat{\bm{\theta}}_1\overset{P}{\rightarrow}
{\bm{\theta}}_0,$ from Theorem 5.4 of~\cite{dzhaparidze1983spectrum}. Thus, with $N_{|\Omega|}$ denoting the cardinality of $\Omega,$
we rewrite the observed Fisher matrix evaluated at a point ${\bm{\theta}}_0'$ squeezed between $\hat{\bm{\theta}}_1$ and the true parameter value as
 \begin{equation*}
 {\bm{F}}\left( {\bm{\theta}}_0'\right)=N_{|\Omega|}
{\bm{\mathcal{F}}}\left({\bm{\theta}}_0\right)
 +N_{|\Omega|}\delta {\bm{ {\mathcal{F}}}}\left({\bm{\theta}}_0\right),
\end{equation*}
for 
\[\delta {\bm{ {\mathcal{F}}}}\left({\bm{\theta}}_0\right)=N_{|\Omega|}^{-1} {\bm{F}}\left( {\bm{\theta}}_0'\right)-{\bm{\mathcal{F}}}\left({\bm{\theta}}_0\right). \]
Using the continuous mapping (or Mann-Wald) theorem, it follows as $\hat{\bm{\theta}}_1\overset{P}{\rightarrow}
{\bm{\theta}}_0,$
\begin{equation*}
\E \|\delta {\bm{ {\mathcal{F}}}}\left({\bm{\theta}}_0\right)\|_F^2=o(1),
\end{equation*}
where $o(1)$ is standard notation for a shrinking quantity. We can therefore deduce 
$\|\delta {\bm{ {\mathcal{F}}}}\left({\bm{\theta}}_0\right)\|_F^2$ is $o_P(1)$.
This is the definition of $\delta {\bm{ {\mathcal{F}}}}\left({\bm{\theta}}_0\right)=o_P(1)$, with the chosen matrix norm being the Frobenius norm.
As $\delta {\bm{ {\mathcal{F}}}}\left({\bm{\theta}}_0\right)$ is of finite size, any sensibly chosen matrix norm, such as the trace norm will be equivalent in order to the Frobenius norm. We therefore may  write
\begin{align*}
N_{|\Omega|}^{-1}{\bm F}\left( {\bm{\theta}}_0'\right)&={\bm{\mathcal{F}}}\left({\bm{\theta}}_0\right)\{\mathbf{I}_{l+p}+o_P(1)\}.
 \end{align*} 
 The same argument holds for the reduced system excluding $\bm{\psi}$. Therefore for ${\bm{\theta}}_0^\ast$ between $\bm\theta_0$ and $\bm{\hat\theta}_0$
 \begin{align*}
N_{|\Omega|}^{-1}\tilde{\bm F}\left({\bm{\theta}}_0^\ast\right)&=\tilde{\bm{\mathcal{F}}}\left({\bm{\theta}}_0\right)
\{\mathbf{I}_l+o_P(1)\}.
 \end{align*}We shall now link the parameter estimates to the score. Write $\nabla=(
{\partial }/{\partial \theta_{11}} \, \dots \, {\partial }/{\partial \theta_{1 l+p}})^T$ for the gradient, and write $\nabla_M=(
{\partial }/{\partial \theta_{01}} \, \dots \, {\partial }/{\partial \theta_{0 l}})^T$ for the gradient under the null model.
We note that performing Taylor series of the score yields
\begin{align*}
\nabla_M \ell_W\left( \hat{\bm{\theta}}_0\right)&=
\nabla_M \ell_W\left( {\bm{\theta}}_0\right)\!-\!\tilde{\bm F}\left( \begin{bmatrix} {\bm{\theta}}_{0M}'' \;\;\bm{\psi}_o\end{bmatrix}^T\right)\!\!
\left\{  \hat{\bm{\theta}}_{0M}- {\bm{\theta}}_{0M}\right\}\\
\nabla \ell_W\left( \hat{\bm{\theta}}_1\right)&=
\nabla \ell_W\left( {\bm{\theta}}_0\right)-{\bm F}\left( {\bm{\theta}}_0^\dagger\right)
\left\{  \hat{\bm{\theta}}_1- {\bm{\theta}}_0\right\},
\end{align*}as we are using a Lagrange form of the Taylor series where the remainder ${\bm{\theta}}_{0M}''$ is squeezed between ${\bm{\theta}}_{0M}$
and $\hat{\bm{\theta}}_{0M}$, as $\hat{\bm{\theta}}_{0M}\overset{P}{\rightarrow}
{\bm{\theta}}_{0M},$ and ${\bm{\theta}}_0^\dagger$ is squeezed between ${\bm{\theta}}_0$
and $\hat{\bm{\theta}}_1$, as $\hat{\bm{\theta}}_1\overset{P}{\rightarrow}
{\bm{\theta}}_0,$ under the null. This naturally produces a representation of 
\begin{equation}
\label{set1}
 \hat{\bm{\theta}}_{0M}- {\bm{\theta}}_{0M}=\left[ \tilde{\cal F}\left({\bm{\theta}}_0\right)+o_P(1)\right]^{-1}N_{|\Omega|}^{-1}\nabla_M \ell_W\left( {\bm{\theta}}_0\right),
\end{equation}
as well as,
\begin{equation}
\label{set2}
 \hat{\bm{\theta}}_{1}- {\bm{\theta}}_{0}=\left[{\cal F}\left({\bm{\theta}}_0\right)+o_P(1)\right]^{-1}N_{|\Omega|}^{-1}\nabla \ell_W\left( {\bm{\theta}}_0\right).
\end{equation}
Calculating the difference between the two estimators for  $\bm{\theta}_M$ from~\eqref{set1} and~\eqref{set2} we get that
\begin{align*}
\begin{pmatrix}\hat{\bm{\theta}}_{1M}- \hat{\bm{\theta}}_{0M}\\
{\mathbf{0}} 
\end{pmatrix} &=\begin{bmatrix}
{\mathbf{I}_l} & {\mathbf{0}} \\
{\mathbf{0}} & {\mathbf{0}}
\end{bmatrix}\left[ {\cal F}+o_P(1)\right]^{-1}N_{|\Omega|}^{-1}\nabla \ell_W\left( {\bm{\theta}}_0\right)\\
&-N_{|\Omega|}^{-1}\begin{bmatrix}
\left\{ \tilde{\cal F}+o_P(1)\right\}^{-1} & {\mathbf{0}} \\
{\mathbf{0}} & {\mathbf{0}}
\end{bmatrix}\nabla \ell_W\left( {\bm{\theta}}_0\right)
\\
&=\left\{ - \begin{bmatrix}
\tilde{\cal F}^{-1} & {\mathbf{0}} \\
{\mathbf{0}} & {\mathbf{0}}
\end{bmatrix}+\begin{bmatrix}
{\mathbf{I}_l} & {\mathbf{0}} \\
{\mathbf{0}} & {\mathbf{0}}
\end{bmatrix} {\cal F}^{-1}\right\}\\
&\quad{\cal F}
\left(  \hat{\bm{\theta}}_1- {\bm{\theta}}_0\right)\left\{\mathbf{I}_{l+p}+o_P(1)\right\}\\
&=\left(\begin{bmatrix}
{\mathbf{I}_l} & {\mathbf{0}} \\
{\mathbf{0}} & {\mathbf{0}}
\end{bmatrix}-
\begin{bmatrix}
{\mathbf{I}_l} & \tilde{\cal F}^{-1}{\cal F}_{\times} \\
{\mathbf{0}} & {\mathbf{0}}
\end{bmatrix}\right)\left\{  \hat{\bm{\theta}}_1- {\bm{\theta}}_0\right\}\\
&\times \left\{\mathbf{I}_{l+p}+o_P(1)\right\}.
\end{align*}
Note that all matrices and submatrices of the Fisher information above should be evaluated at the value ${\bm{\theta}}_0$. This has been omitted for brevity. This expression can then be simplified  to
\begin{equation}\label{bah}\widehat{\bm{\theta}}_{1M}=\widehat{\bm{\theta}}_{0M}-\tilde {\mathcal F}^{-1} {\mathcal F}_{\times}\left(\widehat{\bm{\psi}}-{\bm{\psi}}_o\right)\left\{\mathbf{I}_p+o_P(1)\right\}.\end{equation}

\subsection*{Distribution of the Quadratic Form}
To determine the distribution of the likelihood ratio statistic we shall assume that 1) The process under observation is a Gaussian stationary process possessing a continuous spectrum, 2) ${\bm{\theta}}_0$ lies in an open ball in the parameter set
$\bm{\Theta}$ and 3) the spectrum of the process satisfies certain regularity conditions which will be stated more carefully later in this section. It is not a necessary assumption that the process is Gaussian (see for example work by~\cite{brillinger2001time}), but it is sufficient and easy to state.  
We note that for the example in section VI the null value of the parameter is not in an open set but at the boundary, explaining why we use the parametric bootstrap.

We now return to the quantity $W$ from~\eqref{e:LRstatistic}, and implement an additional Taylor series
for $W$, yet again with the Lagrange form of the remainder (assume ${\bm{\theta}}_0^\ast$ lies in a ball centered at $\hat{\bm{\theta}}_1$ less than $\hat{\bm{\theta}}_0 -\hat{\bm{\theta}}_1$ away from the center), an expansion that is possible because ${\bm{\theta}}_0$ lies in an open ball in the parameter set
$\bm{\Theta}$. Note 
\begin{align}
\nonumber
\ell_W\left(\hat{\bm{\theta}}_0 \right)&=
\ell_W\left(\hat{\bm{\theta}}_1 \right)+\left(\hat{\bm{\theta}}_0-\hat{\bm{\theta}}_1 \right)^T\nabla
\ell_W\left(\hat{\bm{\theta}}_1 \right)\\
&\quad -\frac{1}{2}\cdot \left(\hat{\bm{\theta}}_0-\hat{\bm{\theta}}_1 \right)^T{\bm F}\left( {\bm{\theta}}_0^\ast\right)\left(\hat{\bm{\theta}}_0-\hat{\bm{\theta}}_1 \right).
\end{align}
We can use that $\nabla
\ell_W\left(\hat{\bm{\theta}}_1 \right)=\mathbf{0}$ and then evaluate the variable $W$ defined in~\eqref{e:LRstatistic}
\begin{multline*}
W=2\ell_W\left( \hat{\bm{\theta}}_1\right)-2\ell_W\left( \hat{\bm{\theta}}_0\right)\\=2\ell_W\left(\hat{\bm{\theta}}_1 \right)-2\left[\ell_W\left(\hat{\bm{\theta}}_1 \right)\right.\\
\left.-\frac{1}{2}\cdot \left(\hat{\bm{\theta}}_0-\hat{\bm{\theta}}_1 \right)^T{\bm F}\left( {\bm{\theta}}_0^\ast\right)\left(\hat{\bm{\theta}}_0-\hat{\bm{\theta}}_1 \right) \right]
\\=N_{|\Omega|}\left(  \hat{\bm{\theta}}_0- \hat{\bm{\theta}}_1\right)^T
{\cal F}\left({\bm{\theta}}_0\right)\left(  \hat{\bm{\theta}}_0- \hat{\bm{\theta}}_1 \right)(\mathbf{I}_{l+p}+o_P(1)),
\end{multline*}as ${\bm{\theta}}_0^\ast$ is squeezed between $\hat{\bm{\theta}}_0$ and $\hat{\bm{\theta}}_1$, again using the continuous mapping theorem. Substituting in~\eqref{bah} then yields 
\begin{multline*}
W= N_{|\Omega|}\begin{bmatrix}-\tilde {\mathcal F}^{-1} {\mathcal F}_{\times}\left(\widehat{\bm{\psi}}-{\bm{\psi}}_o\right)\\
\widehat{\bm{\psi}}-{\bm{\psi}}_o
\end{bmatrix}^T{\cal F}\left({\bm{\theta}}_0\right)
\\\times \begin{bmatrix}-\tilde {\mathcal F}^{-1} {\mathcal F}_{\times}\left(\widehat{\bm{\psi}}-{\bm{\psi}}_o\right)\\
\widehat{\bm{\psi}}-{\bm{\psi}}_o
\end{bmatrix}\left\{1+o_P(1)\right\}\\
=N_{|\Omega|}\left(\widehat{\bm{\psi}}-{\bm{\psi}}_o\right)^T\begin{bmatrix} -{\mathcal F}_{\times}^T\tilde {\mathcal F}^{-1} & \mathbf{I}_p
\end{bmatrix}
\begin{bmatrix}
\tilde{\mathcal F} & {\mathcal F}_{\times}\\
{\mathcal F}_{\times}^T&{\mathcal F}_{\circ}
\end{bmatrix} \\ \times
\begin{bmatrix}-\tilde {\mathcal F}^{-1}  {\mathcal F}_{\times}\\ \mathbf{I}_p
\end{bmatrix}
\begin{pmatrix}
\widehat{\bm{\psi}}-{\bm{\psi}}_o
\end{pmatrix}\left\{1+o_P(1)\right\}
\\
=N_{|\Omega|}\left(\widehat{\bm{\psi}}-{\bm{\psi}}_o\right)^T \\ \times \begin{bmatrix} -{\mathcal F}_{\times}^T\tilde {\mathcal F}^{-1}\tilde{\mathcal F} +{\mathcal F}_{\times}^T & -{\mathcal F}_{\times}^T\tilde {\mathcal F}^{-1}{\mathcal F}_\times+{\mathcal F}_{\circ}
\end{bmatrix} \\ \times
\begin{bmatrix}-\tilde {\mathcal F}^{-1}  {\mathcal F}_{\times}\\ \mathbf{I}_p
\end{bmatrix}
\begin{pmatrix}
\widehat{\bm{\psi}}-{\bm{\psi}}_o
\end{pmatrix}\left\{1+o_P(1)\right\}
\\
=N_{|\Omega|}\left(\widehat{\bm{\psi}}-{\bm{\psi}}_o\right)^T\begin{bmatrix}  {\mathcal F}_{\circ}-{\mathcal F}_{\times}^T\tilde {\mathcal F}^{-1}{\mathcal F}_\times
\end{bmatrix}
\begin{pmatrix}
\widehat{\bm{\psi}}-{\bm{\psi}}_o
\end{pmatrix}\\\times\left\{1+o_P(1)\right\}.
\end{multline*}
The final step comes from observing that from~\cite[Theorem 5.5]{dzhaparidze1983spectrum}, writing ${\mathcal{G}}={\mathcal{F}}^{-1}$, and partitioning
the matrix into 
\[{\mathcal{G}}=\begin{bmatrix} {\mathcal{G}}_1 &{\mathcal{G}}_2\\ 
{\mathcal{G}}_2^T & {\mathcal{G}}_3\end{bmatrix},\]
we arrive at 
\begin{equation}
N_{|\Omega|}^{1/2} \left\{\widehat{\bm{\psi}}-{\bm{\psi}}_o\right\}\sim N\left(\bm{0},{\mathcal{G}}_3\right).
\end{equation}
For~\cite[Theorem 5.5]{dzhaparidze1983spectrum} to hold, we need to assume that the observed process is Gaussian, and has a spectrum $S(\omega)$, which for two distinct parameter values are not equal for almost all frequencies, and the inverse spectrum as well as its derivatives with respect to all parameter components are continuous both in frequency and parameter components. Normality can be derived under other assumptions, see e.g.~\cite{brillinger2001time}, but this simplifies the statement of the result.

Note that when giving the normality result from~\cite[Theorem 5.5]{dzhaparidze1983spectrum} we would use ${\bm{\psi}}$ rather than ${\bm{\psi}}_o$, but as the distribution is determined under the assumption that the null holds, we can replace ${\bm{\psi}}$ by ${\bm{\psi}}_o$.
Noting that
\begin{equation}
{\mathcal F}_{\circ}-{\mathcal F}_{\times}^T\tilde {\mathcal F}^{-1}{\mathcal F}_\times 
={\mathcal{G}}_3^{-1},
\end{equation}
we therefore directly arrive at the result by defining the new random vector
\begin{equation}
\bm{Z}_1=N_{|\Omega|}^{1/2} \begin{pmatrix}  {\mathcal F}_{\circ}-{\mathcal F}_{\times}^T\tilde {\mathcal F}^{-1}{\mathcal F}_\times
\end{pmatrix}^{1/2}
\begin{pmatrix}
\widehat{\bm{\psi}}-{\bm{\psi}}_o
\end{pmatrix},
\end{equation}
where we may determine that asymptotically
\[\bm{Z}_1={\mathcal{G}}_3^{-1/2}N_{|\Omega|}^{1/2} \left(\widehat{\bm{\psi}}-{\bm{\psi}}_o\right)\sim N\left(\bm{0},\mathbf{I}_p\right),\]
from which the asymptotic result $\bm{Z}_1^T\bm{Z}_1\sim \chi^2_p$ follows.

\section*{Acknowledgments}
The authors would like to thank the anonymous reviewers for their many important suggestions, and Dr Jorge Ramirez for helpful discussions.

\bibliographystyle{IEEEtran}

\begin{thebibliography}{10}
\providecommand{\url}[1]{#1}
\csname url@samestyle\endcsname
\providecommand{\newblock}{\relax}
\providecommand{\bibinfo}[2]{#2}
\providecommand{\BIBentrySTDinterwordspacing}{\spaceskip=0pt\relax}
\providecommand{\BIBentryALTinterwordstretchfactor}{4}
\providecommand{\BIBentryALTinterwordspacing}{\spaceskip=\fontdimen2\font plus
\BIBentryALTinterwordstretchfactor\fontdimen3\font minus
  \fontdimen4\font\relax}
\providecommand{\BIBforeignlanguage}[2]{{%
\expandafter\ifx\csname l@#1\endcsname\relax
\typeout{** WARNING: IEEEtran.bst: No hyphenation pattern has been}%
\typeout{** loaded for the language `#1'. Using the pattern for}%
\typeout{** the default language instead.}%
\else
\language=\csname l@#1\endcsname
\fi
#2}}
\providecommand{\BIBdecl}{\relax}
\BIBdecl

\bibitem{brands1997radio}
P.~J. Brands, A.~P. Hoeks, L.~A.~F. Ledoux, and R.~S. Reneman, ``A radio
  frequency domain complex cross-correlation model to estimate blood flow
  velocity and tissue motion by means of ultrasound,'' \emph{Ultrasound Med.
  Biol.}, vol.~23, no.~6, pp. 911--920, 1997.

\bibitem{gonella1972rotary}
J.~Gonella, ``A rotary-component method for analysing meteorological and
  oceanographic vector time series,'' \emph{Deep-Sea Res.}, vol.~19, no.~12,
  pp. 833--846, 1972.

\bibitem{lumpkin07}
R.~Lumpkin and M.~Pazos, ``Measuring surface currents with {S}urface {V}elocity
  {P}rogram drifters: the instrument, its data, and some results,'' in
  \emph{{L}agrangian analysis and prediction of coastal and ocean
  dynamics}.\hskip 1em plus 0.5em minus 0.4em\relax Cambridge University Press,
  2007, ch.~2, pp. 39--67.

\bibitem{walker1993complex}
A.~M. Walker, ``Periodogram analysis for complex-valued time series,'' in
  \emph{Developments in Time Series Analysis}, T.~Subba~Rao, Ed.\hskip 1em plus
  0.5em minus 0.4em\relax Chapman and Hall, 1993, pp. 149--163.

\bibitem{mandic2009complex}
D.~P. Mandic and V.~S.~L. Goh, \emph{Complex valued nonlinear adaptive filters:
  noncircularity, widely linear and neural models}.\hskip 1em plus 0.5em minus
  0.4em\relax John Wiley \& Sons, 2009.

\bibitem{schreier2010statistical}
P.~J. Schreier and L.~L. Scharf, \emph{Statistical signal processing of
  complex-valued data: The theory of improper and noncircular signals}.\hskip
  1em plus 0.5em minus 0.4em\relax Cambridge University Press, 2010.

\bibitem{lilly2010bivariate}
J.~M. Lilly and S.~C. Olhede, ``Bivariate instantaneous frequency and
  bandwidth,'' \emph{IEEE T. Signal Proces.}, vol.~58, no.~2, pp. 591--603,
  2010.

\bibitem{walden2013rotary}
A.~T. Walden, ``Rotary components, random ellipses and polarization: a
  statistical perspective,'' \emph{Phil. Trans. R. Soc. A}, vol. 371, 2013.

\bibitem{davenport1961spectrum}
A.~G. Davenport, ``The spectrum of horizontal gustiness near the ground in high
  winds,'' \emph{Q. J. Roy. Meteor. Soc.}, vol.~87, no. 372, pp. 194--211,
  1961.

\bibitem{calman1978interpretation}
J.~Calman, ``On the interpretation of ocean current spectra. part {II}:
  {T}esting dynamical hypotheses,'' \emph{J. Phys. Oceanogr.}, vol.~8, no.~4,
  pp. 644--652, 1978.

\bibitem{picinbono1997second}
B.~Picinbono and P.~Bondon, ``Second-order statistics of complex signals,''
  \emph{IEEE T. Signal Proces.}, vol.~45, no.~2, pp. 411--420, 1997.

\bibitem{rubin2008kinematics}
P.~Rubin-Delanchy and A.~T. Walden, ``Kinematics of complex-valued time
  series,'' \emph{IEEE T. Signal Proces.}, vol.~56, no.~9, pp. 4189--4198,
  2008.

\bibitem{navarro2008arma}
J.~Navarro-Moreno, ``A{RMA} prediction of widely linear systems by using the
  innovations algorithm,'' \emph{IEEE T. Signal Proces.}, vol.~56, no. 7-2, pp.
  3061--3068, 2008.

\bibitem{sykulski2015improper}
A.~M. Sykulski, S.~C. Olhede, and J.~M. Lilly, ``A widely linear complex
  autoregressive process of order one,'' \emph{IEEE T. Signal Proces.},
  vol.~64, no.~23, pp. 6200--6210, 2016.

\bibitem{hamon1974spectral}
B.~V. Hamon and E.~J. Hannan, ``Spectral estimation of time delay for
  dispersive and non-dispersive systems,'' \emph{J. R. Statist. Soc. C},
  vol.~23, no.~2, pp. 134--142, 1974.

\bibitem{robinson1995gaussian}
P.~M. Robinson, ``Gaussian semiparametric estimation of long range
  dependence,'' \emph{Ann. Stat.}, vol.~23, no.~5, pp. 1630--1661, 1995.

\bibitem{gabor1946theory}
D.~Gabor, ``Theory of communication,'' \emph{Proc. IEEE}, vol.~93, no.~26, pp.
  429--457, 1946.

\bibitem{cohen1995time}
L.~Cohen, \emph{Time-frequency analysis}.\hskip 1em plus 0.5em minus
  0.4em\relax Prentice hall, 1995, vol. 778.

\bibitem{gneiting2010matern}
T.~Gneiting, W.~Kleiber, and M.~Schlather, ``Mat{\'e}rn cross-covariance
  functions for multivariate random fields,'' \emph{J. Am. Stat. Soc.}, vol.
  105, no. 491, pp. 1167--1177, 2010.

\bibitem{schreier2008polarization}
P.~J. Schreier, ``Polarization ellipse analysis of nonstationary random
  signals,'' \emph{IEEE T. Signal Proces.}, vol.~56, no.~9, pp. 4330--4339,
  2008.

\bibitem{whittle1953analysis}
P.~Whittle, ``The analysis of multiple stationary time series,'' \emph{J. R.
  Statist. Soc. B}, vol.~15, no.~1, pp. 125--139, 1953.

\bibitem{dzhaparidze1983spectrum}
K.~O. Dzhaparidze and A.~M. Yaglom, ``Spectrum parameter estimation in time
  series analysis,'' in \emph{Developments in Statistics}, P.~R. Krishnaiah,
  Ed.\hskip 1em plus 0.5em minus 0.4em\relax Academic Press, Inc., 1983, pp.
  1--96.

\bibitem{whittle1953estimation}
P.~Whittle, ``Estimation and information in stationary time series,''
  \emph{Ark. Mat.}, vol.~2, no.~5, pp. 423--434, 1953.

\bibitem{dahlhaus1988small}
R.~Dahlhaus, ``Small sample effects in time series analysis: A new asymptotic
  theory and a new estimate,'' \emph{Ann. Stat.}, vol.~16, no.~2, pp. 808--841,
  1988.

\bibitem{marple1999computing}
S.~L. Marple~Jr, ``Computing the discrete-time “analytic” signal via
  {FFT},'' \emph{IEEE T. Signal Proces.}, vol.~47, no.~9, pp. 2600--2603, 1999.

\bibitem{velasco2000whittle}
C.~Velasco and P.~M. Robinson, ``Whittle pseudo-maximum likelihood estimation
  for nonstationary time series,'' \emph{J. Am. Stat. Soc.}, vol.~95, no. 452,
  pp. 1229--1243, 2000.

\bibitem{sykulski2016debiased}
A.~M. Sykulski, S.~C. Olhede, and J.~M. Lilly, ``The de-biased {W}hittle
  likelihood for second-order stationary stochastic processes,'' \emph{arXiv
  preprint:1605.06718}, 2016.

\bibitem{percival1993spectral}
D.~B. Percival and A.~T. Walden, \emph{Spectral Analysis for Physical
  Applications: Multitaper and conventional univariate techniques}.\hskip 1em
  plus 0.5em minus 0.4em\relax Cambridge University Press, 1993.

\bibitem{glover1977adaptive}
J.~R. Glover, ``Adaptive noise canceling applied to sinusoidal interferences,''
  \emph{IEEE T. on Acoust. Speech}, vol.~25, no.~6, pp. 484--491, 1977.

\bibitem{hamon1963estimating}
B.~Hamon and E.~J. Hannan, ``Estimating relations between time series,''
  \emph{J. Geophys. Res.}, vol.~68, no.~21, pp. 6033--6041, 1963.

\bibitem{wahba1980automatic}
G.~Wahba, ``Automatic smoothing of the log periodogram,'' \emph{J. Am. Stat.
  Assoc.}, vol.~75, no. 369, pp. 122--132, 1980.

\bibitem{robinson1994semiparametric}
P.~M. Robinson, ``Semiparametric analysis of long-memory time series,''
  \emph{Ann. Stat.}, vol.~22, no.~1, pp. 515--539, 1994.

\bibitem{lumpkin2010surface}
R.~Lumpkin and S.~Elipot, ``Surface drifter pair spreading in the {N}orth
  {A}tlantic,'' \emph{J. Geophys. Res.-Oceans}, vol. 115, no. C12, p. C12017,
  2010.

\bibitem{sykulski2016lagrangian}
A.~M. Sykulski, S.~C. Olhede, J.~M. Lilly, and E.~Danioux, ``Lagrangian time
  series models for ocean surface drifter trajectories,'' \emph{J. R. Statist.
  Soc. C}, vol.~65, no.~1, pp. 29--50, 2016.

\bibitem{hurvich1998mean}
C.~M. Hurvich, R.~Deo, and J.~Brodsky, ``The mean squared error of {G}eweke and
  {P}orter-{H}udak's estimator of the memory parameter of a long-memory time
  series,'' \emph{J. Time Ser. Anal.}, vol.~19, no.~1, pp. 19--46, 1998.

\bibitem{olhede2009frequency}
S.~C. Olhede, A.~M. Sykulski, and G.~A. Pavliotis, ``Frequency domain
  estimation of integrated volatility for {I}t\^o processes in the presence of
  market-microstructure noise,'' \emph{Multiscale Model. Sim.}, vol.~8, no.~2,
  pp. 393--427, 2009.

\bibitem{cressie1993statistics}
N.~Cressie, \emph{Statistics for Spatial Data}.\hskip 1em plus 0.5em minus
  0.4em\relax Hoboken, New Jersey: John Wiley \& Sons, 1993.

\bibitem{masry1978alias}
E.~Masry, ``Alias-free sampling: {A}n alternative conceptualization and its
  applications,'' \emph{IEEE T. Inform. Theory}, vol.~24, no.~3, pp. 317--324,
  1978.

\bibitem{ollila2004generalized}
E.~Ollila and V.~Koivunen, ``Generalized complex elliptical distributions,'' in
  \emph{Sensor Array and Multichannel Signal Processing Workshop
  Proceedings}.\hskip 1em plus 0.5em minus 0.4em\relax IEEE, 2004, pp.
  460--464.

\bibitem{walden2009testing}
A.~T. Walden and P.~Rubin-Delanchy, ``On testing for impropriety of
  complex-valued {G}aussian vectors,'' \emph{IEEE T. Signal Proces.}, vol.~57,
  no.~3, pp. 825--834, 2009.

\bibitem{chandna2016frequency}
S.~Chandna and A.~T. Walden, ``A frequency domain test for propriety of
  complex-valued vector time series,'' \emph{IEEE T. Signal Proces.}, vol.~65,
  no.~6, pp. 1425--1436, 2016.

\bibitem{tugnait2016testing}
J.~K. Tugnait and S.~A. Bhaskar, ``Testing for impropriety of multivariate
  complex random processes,'' in \emph{IEEE International Conference on
  Acoustics, Speech and Signal Processing (ICASSP)}.\hskip 1em plus 0.5em minus
  0.4em\relax IEEE, 2016, pp. 4264--4268.

\bibitem{vuong1989likelihood}
Q.~H. Vuong, ``Likelihood ratio tests for model selection and non-nested
  hypotheses,'' \emph{Econometrica}, vol.~57, no.~2, pp. 307--333, 1989.

\bibitem{hurvich1989regression}
C.~M. Hurvich and C.~L. Tsai, ``Regression and time series model selection in
  small samples,'' \emph{Biometrika}, vol.~76, no.~2, pp. 297--307, 1989.

\bibitem{vallis2006atmospheric}
G.~K. Vallis, \emph{Atmospheric and oceanic fluid dynamics: Fundamentals and
  large-scale circulation}.\hskip 1em plus 0.5em minus 0.4em\relax Cambridge
  University Press, 2006.

\bibitem{lilly2016fractional}
J.~M. Lilly, A.~M. Sykulski, J.~J. Early, and S.~C. Olhede, ``Fractional
  {B}rownian motion, the {M}at{\'e}rn process, and stochastic modeling of
  turbulent dispersion,'' \emph{arXiv preprint:1605.01684}, 2016.

\bibitem{lacasce2008statistics}
J.~H. LaCasce, ``Statistics from {L}agrangian observations,'' \emph{Prog.
  Oceanogr.}, vol.~77, no.~1, pp. 1--29, 2008.

\bibitem{thomson1982spectrum}
D.~J. Thomson, ``Spectrum estimation and harmonic analysis,'' \emph{Proc.
  IEEE}, vol.~70, no.~9, pp. 1055--1096, 1982.

\bibitem{rhines1975waves}
P.~B. Rhines, ``Waves and turbulence on a beta-plane,'' \emph{J. Fluid. Mech.},
  vol.~69, no.~3, pp. 417--443, 1975.

\bibitem{brillinger2001time}
D.~R. Brillinger, \emph{Time series: Data analysis and theory}.\hskip 1em plus
  0.5em minus 0.4em\relax S{IAM}, 2001.

\end{thebibliography}

\end{document}